\documentclass[prd,nofootinbib,showpacs, 11pt]{revtex4}
\usepackage{graphicx,color}
\begin{document}

\title{ The transition form factors for semi-leptonic weak decays of  $J/\psi$
in QCD sum rules}

\author{Yu-Ming Wang$^{1}$} \author{Hao Zou$^{1}$} \author{Zheng-Tao Wei$^{2}$}
\author{Xue-Qian Li$^{2}$} \author{Cai-Dian L\"{u} $^{1}$}


 \affiliation{$^{1}$Institute of High Energy Physics, P.O. Box
 918(4), Beijing 100049, China}

 \affiliation{$^{2}$Department of Physics, Nankai University, Tianjin
 300071, China}

\vspace*{1.0cm}

\date{\today}
\begin{abstract}

Within the Standard Model, we investigate the semi-leptonic weak
decays of $J/\psi$. The various form factors of $J/\psi$ transiting
to a single charmed meson ($D^{(*)}_{(d,s)}$) are studied in the
framework of the QCD sum rules. These form factors fully determine
the rates of the weak semi-leptonic decays of $J/\psi$ and provide
valuable information about the non-perturbative QCD effects. Our
results indicate that the  decay rate of the semi-leptonic weak
decay mode $J/\psi \to D^{(*)-}_{s}+e^{+}+\nu_{e}$ is at order of
$10^{-10}$.


\end{abstract}

\pacs{13.20.Gd, 13.25.Gv, 11.55.Hx} \maketitle


\section{Introduction}

Although strong and electromagnetic decays of $J/\psi$ have been
extensively studied for several decades, both experimental and
theoretical investigations of weak decays of $J/\psi$ are much
behind. Due to smallness of the strength of weak interaction, the
weak decays of the $J/\psi$ are rare processes. Sanchis-Lonzano
suggested to search for these rare decays whose sum of branching
ratios were estimated to be at the order of $10^{-8}$
\cite{Sanchis-Lonzano}. Such processes hardly drew much attention
because the database was far from reaching such accuracy. Thus, for
a long time, few further researches on this topic were done. Thanks
to the progress of accelerator and detector techniques, more
accurate measurements may be carried out, thus the interest on weak
decays of $J/\psi$ has been revived. The BES collaboration indeed
starts to measure some rare weak decays of $J/\psi$ and eventually
sets an upper bound on the branching ratio of $J/\psi \to
D+e+\nu_{e}$ at order of $10^{-5}$   by using $5.8\times 10^7$
$J/\psi$ database \cite{BES}. The forthcoming upgraded BESIII can
accumulate $10^{10}$ $J/\psi$ per year \cite{BESIII}, which makes it
marginally possible to measure such weak decays of $J/\psi$, at
least one may expect to observe not-null such events. Thus, more
careful theoretical investigation on these decays seems necessary.

Indeed, the weak decays of heavy quarkonium like $J/\psi$ offer an
ideal opportunity of studying non-perturbative QCD effects, because
such systems contain two heavy constituents of the same flavor. The
situation is quite different from that for heavy mesons which
contain only one heavy constituent, and the non-perturbative effects
might be attributed to the light flavor, thus the heavy quark
effective theory (HQET) applies. Moreover, for the weak decay of a
vector meson, the polarization effect may play a role to probe the
underlying dynamics  and hadron structure \cite{Sanchis-Lonzano}.

The weak decay of $J/\psi$ is realized via the spectator mechanism
that the charm quark (antiquark) decays and the antiquark (quark)
acts as a spectator. The characteristic of the decay modes is that
the final state contains a single charmed hadron. The theory of weak
interactions has been thoroughly investigated and the effective
hamiltonian  at the quark level is perfectly formulated. The main
job of calculating the rates of the semi-leptonic decays of $J/\psi$
is to properly evaluate the hadronic matrix elements for $J/\psi\to
D^{(*)}$, namely the transition form factors which are obviously
governed by non-perturbative QCD effects. The main aim of this work
is to calculate the $J/\psi \to D_{(d,s)}^{(*)}$ form factors in the
QCD sum rules.

The weak decay of heavy quarkonium has been studied by virtue of
heavy quark spin symmetry \cite{Sanchis-Lonzano}. In that framework,
the transition form factors of a heavy quarkonium to heavy
pseudoscalar and vector mesons are parameterized by a universal
function $\eta_{12}(v_1 \cdot v_2)$ in analog to the Isgur-Wise
function for the heavy meson transitions. However, the non-recoil
approximation $\eta_{12}(v_1 \cdot v_2)\approx 1$ was used  in
Ref.\cite{Sanchis-Lonzano},  which would  bring up uncontrollable
uncertainties to the estimation of decay widths.  It seems helpful
to re-investigate these processes based on a more rigorous
theoretical framework. Motivated by the arguments,  in this work we
will calculate the form factors for heavy quarkonium $J/\psi$ decays
into a pseudoscalar or vector meson in the QCD sum rules.

As a matter of fact, many authors have tried to evaluate the
transition form factors for the heavy meson and quarkonium system in
various approaches, such as the simple quark model \cite{quark
model}, light-front approach \cite{light front}, the QCD sum rules
\cite{QCDSR 1,QCDSR 2}, the perturbative QCD approach \cite{PQCD}
and etc. The QCD sum-rule approach, which is rooted in the quantum
field theory and fully relativistic, is considered to be one of the
effective tools for analyzing hadronic processes \cite{QCDSR 1}.
Besides evaluation of hadron spectra, the QCD sum-rule technique has
been applied to calculate the pion electromagnetic form factor at
intermediate momentum transfer \cite{ioffe 1, nesterenko}, various
weak decay channels \cite{weak decays of QCDSR 1, weak decays of
QCDSR 2}, the coupling constant of the strong interaction
\cite{coupling constants of the strong interactions} and even to
determine the light cone distribution amplitudes of hadrons
\cite{Chernyak}. The advantage of this method is that the
non-perturbative QCD effects are included in a few parameters such
as the quark- and gluon-condensates which have evident physical
meaning \cite{p. ball}. 

 After this
introduction, we will firstly display the effective Hamiltonian
relevant to the semi-leptonic decays of $J/\psi$ to
$D^{(*)-}_{d(s)}$, and the sum rules for form factors in section
\ref{The standard procedure}. The Wilson coefficients of various
operators which manifest the perturbative QCD effects are also
calculated in this section  with the help of operator product
expansion (OPE) technique.  The numerical analysis on the form
factors are performed in section \ref{Numerical results}. The decay
rates of semi-leptonic decay $J/\psi \to D^{(*)-}_{d(s)} l^+ \nu$
and a comparison of our results with that obtained based on other
approaches are presented in section \ref{decay rate}. In the last
section we draw our conclusion.

\section{$J/\psi \to D^{(*)}_{d(s)}$ transition form factors in the QCD sum rules}
 \label{The standard procedure}

\subsection{Definitions of $J/\psi \to D^{(*)}_{d(s)}$ transition form factors}

For the semi-leptonic decays $J/\psi \to D^{(*)}_{d(s)} l^+\nu_l$,
the effective weak Hamiltonian is given by
\begin{eqnarray}
 \mathcal{H}_{eff}(c \to s(d) l\bar\nu_l)={G_{F} \over \sqrt{2}}V^{*}_{cs(d)}
 \bar{s}(\bar d)\gamma_{\mu}(1-\gamma_5)c \,
 \bar{\nu_l}\gamma^{\mu}(1-\gamma_5)l.
\end{eqnarray}

In order to calculate the rate of a semi-leptonic decay, the
essential ingredient is the hadronic matrix element $\langle
D^{(*)}_{d(s)}|\bar{s}\gamma_{\mu}(1-\gamma_5)c|J/\psi\rangle$ which
is parameterized by various form factors \cite{wirbel}:
\begin{eqnarray}
 &&\langle D_{d(s)}(p_2)|\bar{q}\gamma_{\mu}(1-\gamma_5)c|J/\psi(\epsilon,p_1)\rangle
  \nonumber\\
 &&~~~~~~~=-\epsilon_{\mu\nu\alpha\beta}\epsilon^{\nu}p_1^{\alpha}p_2^{\beta}{2V(q^2)
  \over m_{\psi}+m_{D}}+i(m_{\psi}+m_{D})\left[\epsilon_{\mu}-{\epsilon \cdot q
  \over q^2}q_{\mu}\right]A_1(q^2)\nonumber\\
 &&~~~~~~~~~~+i{\epsilon \cdot q \over m_{\psi}+m_{D}}A_2(q^2)\left[(p_1+p_2)_{\mu}
  -{m_{\psi}^2-m_{D}^2\over q^2} q_{\mu}\right]+
  2i m_{\psi}{\epsilon \cdot q \over q^2}q_{\mu}A_0(q^2), \\
 &&\langle D^{*}_{d(s)}(\epsilon_2,p_2)|\bar{q}\gamma_{\mu}(1-\gamma_5)c|J/\psi
 (\epsilon_1,p_1)\rangle \nonumber\\
 &&~~~~~~~=-i\epsilon_{\mu \nu \alpha \beta}\epsilon_1^{\alpha}\epsilon_2^{*\beta}
  \left[(p_1^{\nu}+p_2^{\nu}-{m_{\psi}^2-m_{D^{*}}^2 \over q^2}q^{\nu})
  \tilde{A}_1(q^2)+{m_{\psi}^2-m_{D^{*}}^2 \over q^2}q^{\nu} \tilde{A}_2(q^2)\right]
  \nonumber \\
 &&~~~~~~~~~~+{i \over m_{\psi}^2-m_{D^{*}}^2}\epsilon_{\mu \nu \alpha \beta}
  p_1^{\alpha} p_2^{\beta} [\tilde{A}_3(q^2) \epsilon_1^{\nu}
  \epsilon_2^{*} \cdot q-\tilde{A}_4(q^2) \epsilon_2^{*\nu} \epsilon_1
  \cdot q] \nonumber\\
 &&~~~~~~~~~~+(\epsilon_1 \cdot \epsilon_2^{*})[-({p_1}_{\mu}+{p_2}_{\mu})
  \tilde{V}_1(q^2)+q_{\mu} \tilde{V}_2(q^2)]\nonumber \\
 &&~~~~~~~~~~+{(\epsilon_1 \cdot q)(\epsilon_2^{*} \cdot q) \over
  m_{\psi}^2-m_{D^{*}}^2}\bigg[({p_1}_{\mu}+{p_2}_{\mu}-{m_{\psi}^2-m_{D^{*}}^2
  \over q^2}q_{\mu}) \tilde{V}_3(q^2)\nonumber \\
 &&~~~~~~~~~~+{m_{\psi}^2-m_{D^{*}}^2 \over q^2}q_{\mu}
  \tilde{V}_4(q^2)\bigg]-(\epsilon_1 \cdot q) {\epsilon_{2}}_{\mu}^{*}
  \tilde{V}_5(q^2) + (\epsilon_{2 }^{*} \cdot q){\epsilon_{1}}_{ \mu}
  \tilde{V}_{6}(q^2), \label{vector vector}
\end{eqnarray}
where the convention ${\rm{Tr}}[\gamma_{\mu} \gamma_{\nu}
\gamma_{\rho} \gamma_{\sigma} \gamma_5]=4 i \epsilon_{\mu \nu \rho
\sigma}$ is adopted. For a transition of $J/\psi$ into a charmed
pseudoscalar meson which is induced by the weak current, there are
four independent form factors: $V,~ A_0,~ A_1,~ A_2$; while there
are ten form factors for $J/\psi$ transiting to a charmed vector
meson which are parameterized as
$\tilde{A}_{i}(i=1,2,3,4),~\tilde{V}_{j}(j=1,2,3,4,5,6)$. It is
worthwhile to emphasize that the parametrization of the hadronic
matrix element for $J/\psi$ to vector meson given in Eq.
(\ref{vector vector}) is less studied before. Similar  matrix
element for a transition of a vector  to another vector meson which
is induced by the electromagnetic current was investigated by Kagan
in Ref. \cite{kagan}.

\subsection{The transition form factors in the QCD sum rules}

In this subsection, we calculate transition form factors of $J/\psi
\to D_{(d,s)}^{(*)-}$  in the QCD sum rules. Here we present the
formulations for $J/\psi \to D_s^{(*)-}$ transition explicitly,
while the expressions for $J/\psi \to D^{(*)-}$ can be obtained by
simple replacements of $D_s^{(*)-} \to D^{(*)-}$ and $s$ quark to
$d$ quark.

\subsubsection{The matrix element for $J/\psi\to D^{-}_{s}$}

Following the standard procedure of the QCD sum rules \cite{ioffe
1}, we write the three-point correlation function for $J/\psi$ to
$D^{-}_{s}$ as
\begin{eqnarray} \label{cf1}
\Pi_{\mu \nu}=i^2 \int d^4 x d^4 y e^{-i p_1 \cdot y +i p_2 \cdot
x}\langle 0 |j_5^{D_s^-}(x) j_{\mu}(0)j_{\nu}^{J/\psi}(y) | 0
\rangle, \label{correlator pseudoscalar}
\end{eqnarray}
where the current $j_{\nu}^{J/\psi}(y)=\bar{c}(y) \gamma _{\nu}
c(y)$ represents the $J/\psi$ channel;  $j_{\mu}(0)=\bar{s} \gamma
_{\mu}(1-\gamma_5) c $ is the weak current and
$j_{5}^{D^{-}}(x)=\bar{c}(x) i \gamma _{5 } s(x)$ corresponds to the
$D_s^{-}$ channel. In terms of the following definitions,
\begin{eqnarray}
\langle 0|\bar{c} \gamma_{\nu}
c|J/\psi\rangle=m_{\psi}f_{\psi}\epsilon_{\nu}^{\lambda},\qquad
\langle 0|\bar{c} i \gamma_{5 } s|D_s\rangle={f_{D_s} m_{D_s}^2
\over {m_c+m_s}},
\end{eqnarray}
we can insert a complete set of hadronic states with the quantum
numbers the same as $J/\psi$ and $D_s^-$ to achieve the hadronic
representation of the correlator (\ref{correlator pseudoscalar})
\begin{eqnarray}
\Pi_{\mu \nu}={f_{D_s} m_{D_s}^2\langle D_s|j_{\mu}|J/\psi\rangle
m_{\psi} f_{\psi}\epsilon_{\nu}^{*\lambda} \over
(m_{J/\psi}^2-p_1^2)(m_{D_s}^2-p_2^2)(m_c+m_s)}+
\mathrm{contributions}\,\,\, \mathrm{from}\,\,\,
\mathrm{higher}\,\,\, \mathrm{states}.
\end{eqnarray}
Obviously the concerned lowest hadronic states are $J/\psi$ and
$D_s$, while the terms with ``higher states'' represent
contributions coming from higher excited states and continuum. Using
the double dispersion relation, the contributions of excited states
and continuum can be expressed as
\begin{eqnarray}
\mathrm{contributions}\,\,\, \mathrm{from}\,\,\,
\mathrm{higher}\,\,\, \mathrm{states}=\int\int_{\sum_{12}} ds_1 ds_2
{\rho^{h}_{\mu \nu }(s_1,s_2,q^2) \over
(s_1-p_1^2)(s_2-p_2^2)}+\mathrm{subtraction} \,\,\, \mathrm{terms},
\end{eqnarray}
where $\sum_{12}$ denotes the integration region in the $(s_1,s_2)$
plane. $\rho^{h}_{\mu \nu }$ is the spectral density at the hadron
level. The subtraction terms are polynomials of either $p_1$ or
$p_2$, which should disappear after performing the double Borel
transformation $\hat{\mathcal{B}}_{M_1^2}
\hat{\mathcal{B}}_{M_2^2}$, with
\begin{eqnarray}
\hat{\mathcal{B}}_{M_i^2}=\lim_{\stackrel{-p_i^2,n \to
\infty}{-p_i^2/n=M^2}} \frac{(-p_i^2)^{(n+1)}}{n!}\left(
\frac{d}{dp_i^2}\right)^n.
\end{eqnarray}

On the other side, we calculate the correlation function at the
quark level by using the OPE as
\begin{eqnarray}
\Pi_{\mu \nu}=-f_{0} \epsilon_{\mu \nu \alpha \beta
}p_1^{\alpha}p_2^{\beta}-i(f_1 {p_1}_{\mu} {p_{1}}_{\nu} +f_2
{p_2}_{\mu} {p_{2}}_{\nu} +f_3 {p_2}_{\mu} {p_{1}}_{\nu}+f_4
{p_1}_{\mu} {p_{2}}_{\nu}+f_5 g_{\mu \nu}),
\end{eqnarray}
where each coefficient contains contributions from both
perturbative part and the non-perturbative part whose effects
manifest in several typical condensates,
\begin{eqnarray}
f_i=f_i^{pert} {\mathbf{I}} + f_{i}^{qq} \langle \bar{q} q\rangle +
f_{i}^{GG} \langle G G \rangle + f_{i}^{qGq} \langle \bar{q} G
q\rangle +..., \label{fi expansion}
\end{eqnarray}
with $f_i^{pert}, f_{i}^{qq}, f_{i}^{GG}, f_{i}^{qGq},...$ denoting
the contributions to the correlation functions from dimension 0,\,
3, \, 4, \, 5,... operators. By the quark-hadron duality, one may
match the two different representations of the correlation function
and perform the double Borel transformation on variables $p_1$ and
$p_2$, then we get the sum rules for the form factors
\begin{eqnarray}
\label{V pseudoscalar}V(q^2)&=&-{(m_c+m_s)(m_{\psi}+m_{D_s})\over
2m_{\psi}f_{\psi}f_{D_s}m_{D_s}^2}e^{m_{\psi}^2 / M_1^2}e^{m_{D_s}^2
/ M_2^2}M_1^2 M_2^2\hat{\mathcal{B}}f_0,
\\
A_1(q^2)&=&{(m_c+m_s)\over(m_{\psi}+m_{D_s})
m_{\psi}f_{\psi}f_{D_s}m_{D_s}^2}e^{m_{\psi}^2 / M_1^2}e^{m_{D_s}^2
/ M_2^2}M_1^2 M_2^2\hat{\mathcal{B}}f_5,
\\
A_2(q^2)&=&-{(m_c+m_s)(m_{\psi}+m_{D_s})\over
2m_{\psi}f_{\psi}f_{D_s}m_{D_s}^2}e^{m_{\psi}^2 / M_1^2}e^{m_{D_s}^2
/ M_2^2}M_1^2 M_2^2\hat{\mathcal{B}}(f_2+f_4),
\\
A_0(q^2)&=&-{(m_c+m_s)\over
2m_{\psi}^2f_{\psi}f_{D_s}m_{D_s}^2}e^{m_{\psi}^2 /
M_1^2}e^{m_{D_s}^2 / M_2^2}M_1^2
M_2^2[\hat{\mathcal{B}}(f_2+f_4){m_{\psi}^2-m_{D_s}^2 \over
2}-\hat{\mathcal{B}}(f_2-f_4) {q^2 \over 2}-\hat{\mathcal{B}} f_5].
\label{A0 pseudoscalar}
\end{eqnarray}

\subsubsection{The matrix element for $J/\psi \to D^{*-}_{s}$}
\label{sum rules 2}

The three-point correlation function of $J/\psi$ to $D^{*-}_{s}$
is
\begin{eqnarray}
\Pi_{\mu \nu \rho}=i^2 \int d^4 x d^4 y e^{-i p_1 \cdot y +i p_2
\cdot x}\langle 0 |j_{\rho}^{D^{*}_s}(x)
j_{\mu}(0)j_{\nu}^{J/\psi}(y) | 0 \rangle,
\end{eqnarray}
where the current
$j_{\rho}^{D^{*}_s}(x)=\bar{c}(x)\gamma_{\rho}s(x)$ denotes the
$D^{*-}_s$ channel, and $j_{\nu}^{J/\psi}(y)$, $j_{\mu}(0)$ are
defined as in the above subsection. One the one hand, inserting the
hadron states, the correlation function is written as
\begin{eqnarray}
\Pi_{\mu \nu \rho}={m_{D^{*}_s} {f_{D^{*}_s}}
{\epsilon'_{\rho}}^{\lambda'}\langle D^{*}_s|j_{\mu}|J/\psi\rangle
m_{J/\psi} f_{J/\psi}\epsilon_{\nu}^{*\lambda} \over
(m_{J/\psi}^2-p_1^2)(m_{D^{*}_s}^2-p_2^2)}+ \int\int ds_1 ds_2
{\rho^{h}_{\mu \nu \rho}(s_1,s_2,q^2) \over
(s_1-p_1^2)(s_2-p_2^2)}+\mathrm{subtraction} \,\,\,
\mathrm{terms}.
\end{eqnarray}
On the other hand, the correlation function at the quark level is
formulated as
\begin{eqnarray}
 &&\Pi_{\mu \nu \rho}=i F_{1} \epsilon_{\mu \nu \alpha \beta}
  p_1^{\alpha}p_2^{\beta}{p_1}_{\rho}+i F_{2} \epsilon_{\mu \nu
  \alpha \beta }p_1^{\alpha}p_2^{\beta}{p_2}_{\rho}+i F_{3}
  \epsilon_{\mu \rho \alpha \beta}p_1^{\alpha}p_2^{\beta}{p_1}_{\nu}
  +iF_{4} \epsilon_{\mu \rho \alpha
  \beta}p_1^{\alpha}p_2^{\beta}{p_2}_{\nu} \nonumber \\
 &&~~+i F_{5} \epsilon_{\nu \rho\alpha \beta}p_1^{\alpha}p_2^{\beta}{p_1}_{\mu}
  +i F_{6} \epsilon_{\nu \rho \alpha \beta}p_1^{\alpha}p_2^{\beta}{p_2}_{\mu}
  +F_{7}g_{\mu \nu}{p_1}_{\rho}+F_{8}g_{\mu \rho}{p_1}_{\nu}+
  F_{9}g_{\nu\rho}{p_1}_{\mu}+F_{10}g_{\mu \nu}{p_2}_{\rho}
  \nonumber \\
 &&~~+F_{11}g_{\mu \rho}{p_2}_{\nu}+F_{12}g_{\nu \rho}{p_2}_{\mu}
  +F_{13}{p_1}_{\mu}{p_1}_{\nu}{p_1}_{\rho}
  +F_{14}{p_2}_{\mu}{p_2}_{\nu}{p_1}_{\rho}+F_{15}{p_1}_{\mu}{p_2}_{\nu}{p_1}_{\rho}
  +F_{16}{p_2}_{\mu}{p_1}_{\nu}{p_1}_{\rho}\nonumber \\
 &&~~+F_{17}{p_2}_{\mu}{p_2}_{\nu}{p_2}_{\rho}+F_{18}{p_1}_{\mu}{p_1}_{\nu}{p_2}_{\rho}
  +F_{19}{p_2}_{\mu}{p_1}_{\nu}{p_2}_{\rho}+F_{20}{p_1}_{\mu}{p_2}_{\nu}{p_1}_{\rho},
\end{eqnarray}
where each coefficient $F_{i}$ includes contributions from both
perturbative and nonperturbative parts, and is written explicitly
as
\begin{eqnarray}
F_i=F_i^{pert} {\mathbf{I}} + F_{i}^{qq} \langle \bar{q} q\rangle
+ F_{i}^{GG} \langle G G \rangle + F_{i}^{qGq} \langle \bar{q} G
q\rangle +.... \label{Fi expansion}
\end{eqnarray}
Again, equating the correlation functions calculated in these two
frameworks and performing the Borel transformations on both sides,
we derive the form factors of $J/\psi\to D^{*-}_{s}$ as
\begin{eqnarray}
\label{A1 vector}\tilde{A}_1(q^2)&=&-{1 \over 4
m_{\psi}f_{\psi}m_{D^{*-}}f_{D^{*}_s}}e^{m_{\psi}^2 /
M_1^2}e^{m_{D^{*}_s}^2 / M_2^2}M_1^2 M_2^2
\hat{\mathcal{B}}[(F_5-F_6) q^2
+(F_5+F_6)(m_{\psi}^2-m_{D^{*}_s}^2)],\\
 \tilde{A}_2(q^2)&=&{m_{D^{*}_s}^4-2(q^2+m_{\psi}^2)m_{D^{*}_s}^2
 +(q^2-m_{\psi}^2)^2\over 4(m_{D^{*}_s}^2-m_{\psi}^2)
 m_{\psi}f_{\psi}m_{D^{*}_s}f_{D^{*}_s}}e^{m_{\psi}^2 /
 M_1^2}e^{m_{D^{*}_s}^2 / M_2^2}M_1^2 M_2^2 \hat{\mathcal{B}}(F_5+F_6),
\\
\tilde{A}_3(q^2)&=&{m_{\psi}^2-m_{D^{*}_s}^2 \over
m_{\psi}f_{\psi}m_{D^{*}_s}f_{D^{*}_s}}e^{m_{\psi}^2 /
M_1^2}e^{m_{D^{*}_s}^2 / M_2^2}M_1^2 M_2^2
\hat{\mathcal{B}}(F_1-F_5),
\\
\tilde{A}_4(q^2)&=&{{m_{\psi}}^2-m_{D^{*}_s}^2\over
m_{\psi}f_{\psi}m_{D^{*}_s}f_{D^{*}_s}}e^{m_{\psi}^2 /
M_1^2}e^{m_{D^{*}_s}^2 / M_2^2}M_1^2
M_2^2\hat{\mathcal{B}}(F_4+F_6),
\\
\tilde{V}_1(q^2)&=&-{1 \over 2
m_{\psi}f_{\psi}m_{D^{*}_s}f_{D^{*}_s}}e^{m_{\psi}^2 /
M_1^2}e^{m_{D^{*}_s}^2 / M_2^2}M_1^2
M_2^2\hat{\mathcal{B}}(F_{9}+F_{12}),
\\
\tilde{V}_2(q^2)&=&{1 \over 2
m_{\psi}f_{\psi}m_{D^{*}_s}f_{D^{*}_s}}e^{m_{\psi}^2 /
M_1^2}e^{m_{D^{*}_s}^2 / M_2^2}M_1^2
M_2^2\hat{\mathcal{B}}(F_{9}-F_{12}),
\\
\tilde{V}_{3}(q^2)&=&{m_{D^{*}_s}^2 - {m_{\psi}}^2 \over 2
m_{\psi}f_{\psi}m_{D^{*}_s}f_{D^{*}_s}}e^{m_{\psi}^2 /
M_1^2}e^{m_{D^{*}_s}^2 / M_2^2}M_1^2
M_2^2\hat{\mathcal{B}}(F_{14}+F_{15}),
\\
\tilde{V}_{4}(q^2)&=&{1\over 2
m_{\psi}f_{\psi}m_{D^{*}_s}f_{D^{*}_s}}e^{m_{\psi}^2 /
M_1^2}e^{m_{D^{*}_s}^2 / M_2^2}M_1^2 M_2^2
\hat{\mathcal{B}}[(F_{14}-F_{15})q^2+(F_{14}+F_{15}) (m_{D^{*}_s}^2
- {m_{\psi}}^2) ],
\\
\tilde{V}_5(q^2)&=&{1 \over
m_{\psi}f_{\psi}m_{D^{*}_s}f_{D^{*}_s}}e^{m_{\psi}^2 /
M_1^2}e^{m_{D^{*}_s}^2 / M_2^2}M_1^2 M_2^2\hat{\mathcal{B}}F_{11},
\\
\tilde{V}_6(q^2)&=&{1 \over
m_{\psi}f_{\psi}m_{D^{*-}}f_{D^{*}_s}}e^{m_{\psi}^2 /
M_1^2}e^{m_{D^{*}_s}^2 / M_2^2}M_1^2 M_2^2\hat{\mathcal{B}}F_{7}.
\label{V6 vector}
\end{eqnarray}

\subsection{The Wilson coefficients for correlation function $\Pi_{\mu\nu}$}
 \label{Wilson coefficients}

In this subsection we calculate the Wilson coefficients which are
defined above. To guarantee sufficient theoretical accuracy, the
correlation functions are required to be expanded up to
dimension-5 operators, namely quark-gluon mixing condensate. The
dimension-6 operators, such as the four quark condensates, are
small and further suppressed by $O({\alpha}_s^2)$, so can be
safely neglected in our calculations.

The diagrams which depict the contributions from the perturbative
part and nonperturbative condensates are shown in Fig.~\ref{wilson
coefficients graph}. The first diagram results in the Wilson
coefficient of the unit operator; the second diagram is relevant to
the contribution of quark condensate, where the heavy-quark
condensate is neglected. The Wilson coefficient of the two-gluon
condensate operator is obtained from Fig.~\ref{wilson coefficients
graph}(c-h). The last two diagrams Fig.~\ref{wilson coefficients
graph}(i-j) stand for the contribution of quark-gluon mixing
condensate. In this work, all of the Wilson coefficients are
calculated at the lowest order in the running coupling constant of
strong interaction.

\begin{figure}[tb]
\begin{center}
\begin{tabular}{ccc}
\includegraphics[scale=0.8]{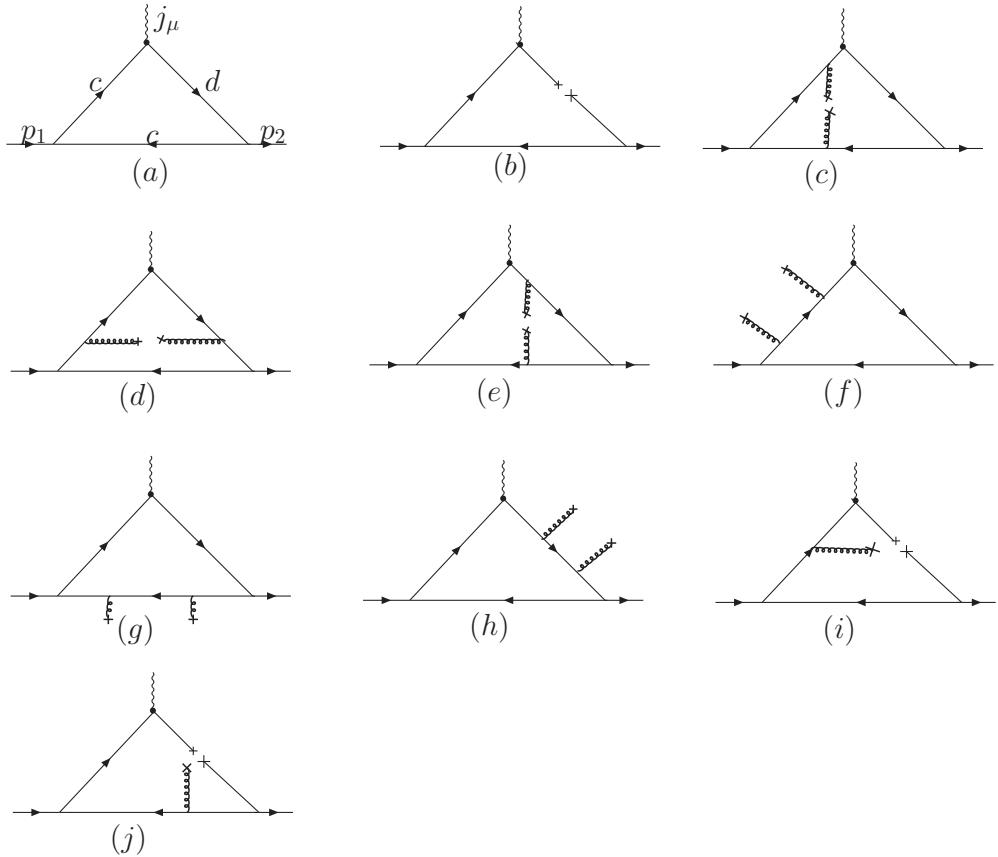}
\vspace{-6 cm}
\end{tabular}
\caption{Graphs for the Wilson coefficients in the operator product
expansion of the correlation function. (a) is for the contribution
of unit operator; (b) for the two-quark condensate; (c-h) describe
the contributions from gluon condensate, (i-j) is for the
quark-gluon mixing condensate.} \label{wilson coefficients graph}
\end{center}
\end{figure}

\subsubsection{Perturbative contributions to Wilson coefficients for
 $\Pi_{\mu \nu}$ }

The perturbative contribution to the three-point correlation
function  $\Pi_{\mu \nu}$ shown in Fig. \ref{wilson coefficients
graph} (a) is included in the following amplitude
\begin{eqnarray}
C^{pert}_{\mu \nu}=3 i^2 \int {d^4 k \over (2 \pi)^4} (-1)
{\rm{Tr}}[\gamma_{\nu}{i \over \not k -m_c } i \gamma_5 {i \over
\not {p_2} + \not k -m_q} \gamma_{\mu}(1-\gamma_5) {i \over \not
{p_1} +\not k -m_c}], \label{perturbative contribution 1}
\end{eqnarray}
where $m_q$ denotes the mass of the light quark in the $D$ meson,
and the factor ``3" is due to the color loop. Using the dispersion
relation, $C^{pert}_{0\mu \nu}$ is written as
\begin{eqnarray}
C^{pert}_{0\mu \nu}=\int\int ds_1 ds_2 {\rho^{pert} _{\mu \nu}
(s_1,s_2,q^2) \over (s_1-p_1^2) (s_2-p_2^2)}. \label{C0}
\end{eqnarray}
The integration region is determined by the following condition
\begin{eqnarray}
-1 \leq  {2 s_1 (s_2+m_c^2-m_q^2)-s_1 (s_1+s_2 -q^2) \over
\lambda^{1/2}(s_1,s_2,q^2) \lambda^{1/2}(m_c^2,s_1,m_c^2) }\leq 1,
\label{integral region}
\end{eqnarray}
where $\lambda(a,b,c)= a^2+b^2+c^2 -2ab-2ac-2bc $. The standard way
to calculate the spectral function $\rho _{\mu \nu} (s_1,s_2,q^2)$
is described below \cite{ioffe 1}: Firstly, it is essential to
calculate the double discontinuity of the amplitude, which can be
realized by putting all the  internal quark lines  of
Fig.~\ref{wilson coefficients graph} (a) on their mass-shell and
substituting the denominators of the quark propagators by the
$\delta$ functions based on the Cutkosky's cutting rule,
\begin{eqnarray}
{1 \over k^2-m^2 +i \epsilon } \rightarrow -2 \pi i \delta (k^2-m^2).
\end{eqnarray}
Then, the spectral function can be easily achieved.  Finally, we
get the expression of the spectral function in the following form
\begin{eqnarray}
\rho _{\mu \nu} (p_1^2,p_2^2,q^2)&=& {3 \over (2 \pi i)^2}  (- 2 \pi
i)^3 \int {d^4 k \over (2 \pi)^4} {\rm{Tr}}[\gamma_{\nu} (\not k +
m_c ) \gamma_5 (\not {p_2} + \not k + m_q)
\gamma_{\mu}(1-\gamma_5) (\not {p_1} +\not k +m_c)] \nonumber \\
&&\delta(k^2-m_c^2)  \delta[(p_2+k)^2-m_q^2]
\delta[(p_1-k)^2-m_c^2].
\end{eqnarray}

After tedious calculations, one finally obtains the perturbative
contribution to the correlation function, which can be decomposed as
the sum of various terms according to different Lorentz structures,
namely,
\begin{eqnarray}
\rho^{pert}_{\mu \nu}=-{\rho}^{pert}_{0} \epsilon_{\mu \nu \alpha
\beta }p_1^{\alpha}p_2^{\beta}-i({\rho}^{pert}_1 {p_1}_{\mu}
{p_{1}}_{\nu} +{\rho}^{pert}_2 {p_2}_{\mu} {p_{2}}_{\nu}
+{\rho}^{pert}_3 {p_2}_{\mu} {p_{1}}_{\nu}+{\rho}^{pert}_4
{p_1}_{\mu} {p_{2}}_{\nu}+{\rho}^{pert}_5 g_{\mu \nu}).
\label{spectral density 1}
\end{eqnarray}
The expressions for the ${\rho}^{pert}_{i}$ are a bit more
tedious, so that we will display their explicit forms in Appendix
\ref{Wilson coefficients for D}.

\subsubsection{The quark condensate contribution}

Now we turn to calculating the Wilson coefficient of the quark
condensate operator, which are shown in Fig. \ref{wilson
coefficients graph} (b).
 One can easily find that
it does not contribute to the correlation function after performing
the double Borel transformation on both variables $p_1^2$ and
$p_2^2$, since the propagator of this diagram  ${1 \over
(p_2^2-m_c^2) (q^2-m_c^2)}$   only depends on variable $p_2^2$. In
other words, the Wilson coefficient of dimension-3 two quark
condensate turns to be zero in the leading order of heavy quark mass
expansion after carrying out the double Borel transformation. As can
be seen, vanishing of the contributions from quark condensate is
independent on the structures of the effective vertices, therefore,
it also does not contribute to the decays of $J/\psi$ into a vector
meson for the same reason. Below, we do not need to investigate the
contributions of quark condensate to $J/\psi \to D^{*}$ based on
this argument.

\subsubsection{The contribution from gluon condensate}

The diagrams which determine the Wilson coefficient of the gluon
condensate are shown in  Fig.~\ref{wilson coefficients graph} (c-h).
The standard way  is using the so-called fixed -point gauge
technique. The gauge fixing condition is
\begin{eqnarray}
x^{\mu} A^{a}_{\mu} =0,
\end{eqnarray}
where $A^{a}_{\mu}$ is the gluon field. In the momentum space,
$A^a_{\mu}(k)$ is transformed to the gauge invariant field
strength as
\begin{eqnarray}
A^{a}_{\mu}(k)= -{i \over 2} (2 \pi)^4 G ^{a}_{ \rho \mu} (0)
{\partial \over \partial k_{\rho} } \delta^4(k) +...
\end{eqnarray}

Indeed, the loop integral
\begin{eqnarray}
I_{\mu_1, \mu_2,... \mu_n}(a,b,c)=\int {d^4 k \over (2 \pi)^4}
{k_{\mu_1} k_{\mu_2}... k_{\mu_n} \over [k^2-m^2]^a
[(p_1+k)^2-m_1^2]^b [(p_2+k)^2-m_2^2]^c},
\end{eqnarray}
which is encountered in the work, is not easy to be performed by the
Feynman parameter method. One alternative way to calculate this kind
of integrals has been extensively discussed in the Ref. \cite{p.
ball, kiselev, Coulomb corrections, t.m. aliev}, where the authors
suggested to work in the Euclidean space-time and employ the
Schwinger representation for propagators. Instead, in our work, we
follow the method employed in Ref. \cite{yangmz, yangmz more},
namely, directly calculate the imaginary part of the integrals in
terms of the Cutkosky's rule.

With the help of the Mathematical package ``FeynCalc'', we finally
get the contributions of Fig. \ref{wilson coefficients graph} (c-h)
at the price of  some long and tedious derivations and
time-consuming computer computations. The contributions of the
gluon-condensates from various sources cancel each other completely
after carrying out the double Borel transformation to the variables
$p_1^2$ and $p_2^2$. Therefore the diagrams  involving the gluon
condensate do not contribute to the transition of vector meson
$J/\psi$ to a pseudoscalar $D$ meson. This argument also applies to
the transition of a pseudoscalar meson to a vector which was
discussed in Ref. \cite{yangmz, yangmz more}, since topologies of
the Feynman diagrams which result in the Wilson coefficient of the
gluon condensate are the same. As analyzed later, it is also true
for the transition of $J/\psi$ to a vector meson. However, we find
that the flavor-changing neutral current process can receive
non-zero contributions from the gluon condensate. It should be noted
that the null contributions of gluon condensates to sum rules for
the weak transition $c\to s(d)$ are different from that obtained in
Ref. \cite{kiselev, Coulomb corrections, t.m. aliev}, where the
method they adopted  does not allow for the substraction of
continuum contributions.

\subsubsection{The quark-gluon mixing condensate contribution}

Finally, we go on calculating the Wilson coefficients of  the
dimension-5 operator $\langle \bar{q}Gq \rangle$. Only two diagrams
shown in Fig. \ref{wilson coefficients graph} (i-j) are involved.
Concentrating on these two diagrams, we  find that they do not
contribute to the correlation function, due to  the same reason
 as that for the null contribution from quark condensate, namely,
only the variable $p_2^2$ appears in the propagators, the amplitude
will vanish due to the double Borel transformation.

As mentioned at the beginning of this section, we do not consider
the four quark condensate, hence only the perturbative part which
corresponds to Fig.1 (a), offers a non-zero contribution to the
correlation function.

\subsection{The Wilson coefficients for the operators contributing to the
correlation function $\Pi_{\mu \nu \rho}$}
\label{Wilson coefficients 2}

After above lengthy discussions, a computation of the correlation
function $\Pi_{\mu \nu \rho}$ which determines the transition
amplitude of $J/\psi$ to a vector meson is straightforward.
Repeating the previous calculations but replacing the vertex for
the pseudoscalar meson to that for a vector meson, one can obtain
the expressions of the Wilson coefficients for all the concerned
operators.

\subsubsection{The calculations of the perturbative contribution to
 $\Pi_{\mu \nu \rho}$}

The Wilson coefficient of the perturbative part corresponding to
Fig. \ref{wilson coefficients graph} (a) is
\begin{eqnarray}
C^{pert}_{\mu \nu \rho}=3 i^2 \int {d^4 k \over (2 \pi)^4} (-1)
{\rm{Tr}}[\gamma_{\nu}{i \over \not k -m_c } \gamma_{\rho} {i \over
\not {p_2} + \not k -m_d} \gamma_{\mu}(1-\gamma_5) {i \over \not
{p_1} +\not k -m_c}].
\end{eqnarray}
We rewrite it in the form of dispersion integrals for the sake of
connecting it to the hadronic spectral density based on the
assumption  of the quark-hadron duality, as
\begin{eqnarray}
C^{pert}_{\mu \nu \rho}=\int\int ds_1 ds_2 {{\rho}^{pert} _{\mu \nu
\rho} (s_1,s_2,q^2) \over (s_1-p_1^2) (s_2-p_2^2)}.
\end{eqnarray}
The integration region is the same as that for the $C^{pert}_{\mu
\nu}$, which is presented in Eq. (\ref{integral region}).
Setting all the internal quark lines on their mass shells,  we
derive the spectral function ${\rho}^{pert} _{\mu \nu \rho}$ as
\begin{eqnarray}
{\rho}^{pert} _{\mu \nu \rho}&=&i {\rho'}^{pert}_{1} \epsilon_{\mu
  \nu \alpha \beta} p_1^{\alpha}p_2^{\beta}{p_1}_{\rho}+i
  {\rho'}^{pert}_{2} \epsilon_{\mu \nu \alpha \beta}
  p_1^{\alpha}p_2^{\beta}{p_2}_{\rho}+i {\rho'}^{pert}_{3}
  \epsilon_{\mu \rho \alpha \beta}p_1^{\alpha}p_2^{\beta}{p_1}_{\nu}
  +i {\rho'}^{pert}_{4} \epsilon_{\mu \rho \alpha \beta}
  p_1^{\alpha}p_2^{\beta}{p_2}_{\nu} \nonumber \\
 &&+i {\rho'}^{pert}_{5}\epsilon_{\nu \rho \alpha \beta}
  p_1^{\alpha}p_2^{\beta}{p_1}_{\mu} +i{\rho'}^{pert}_{6}\epsilon_{\nu\rho\alpha\beta}
  p_1^{\alpha}p_2^{\beta}{p_2}_{\mu}+{\rho'}^{pert}_{7}g_{\mu \nu}
  {p_1}_{\rho}+{\rho'}^{pert}_{8}g_{\mu \rho}{p_1}_{\nu}
  +{\rho'}^{pert}_{9}g_{\nu \rho}{p_1}_{\mu} \nonumber \\
 &&+{\rho'}^{pert}_{10}g_{\mu \nu}{p_2}_{\rho}
  +{\rho'}^{pert}_{11}g_{\mu \rho}{p_2}_{\nu}+{\rho'}^{pert}_{12}g_{\nu
  \rho}{p_2}_{\mu}+{\rho'}^{pert}_{13}{p_1}_{\mu}{p_1}_{\nu}{p_1}_{\rho}
  +{\rho'}^{pert}_{14}{p_2}_{\mu}{p_2}_{\nu}{p_1}_{\rho}
  +{\rho'}^{pert}_{15}{p_1}_{\mu}{p_1}_{\nu}{p_1}_{\rho} \nonumber \\
  && +{\rho'}^{pert}_{16}{p_2}_{\mu}{p_1}_{\nu}{p_1}_{\rho}
  +{\rho'}^{pert}_{17}{p_2}_{\mu}{p_2}_{\nu}{p_2}_{\rho}
  +{\rho'}^{pert}_{18}{p_1}_{\mu}{p_1}_{\nu}{p_2}_{\rho}
  +{\rho'}^{pert}_{19}{p_2}_{\mu}{p_1}_{\nu}{p_2}_{\rho}
  +{\rho'}^{pert}_{20}{p_1}_{\mu}{p_2}_{\nu}{p_1}_{\rho}.
\end{eqnarray}
Only ${\rho'}^{pert}_{i}\,\,(i=1,4,5,6,7,9,11,12,14,15)$ are related
to the form factors $\tilde{V}_1$, $\tilde{V}_2$, $\tilde{V}_3$,
$\tilde{V}_4$, $\tilde{V}_5$, $\tilde{V}_6$, $\tilde{A}_1$,
$\tilde{A}_2$, $\tilde{A}_3$, $\tilde{A}_4$, and we display their
expressions in Appendix \ref{Wilson coefficients for Dstar}.

\subsubsection{The contribution of gluon condensate to $\Pi_{\mu \nu \rho}$}

Similar to the derivation made above, we easily obtain the Wilson
coefficient of the gluon condensate which may contribute to the
correlation function $\Pi_{\mu \nu \rho}$.  Then we rewrite the
Wilson coefficient in the form of dispersion integrals:
\begin{eqnarray}
C^{GG}_{\mu \nu \rho}=\int\int ds_1 ds_2 {\rho^{GG} _{\mu \nu \rho}
(s_1,s_2,q^2) \over (s_1-p_1^2) (s_2-p_2^2)}, \label{C4}
\end{eqnarray}
where the integral region is the same as that for the perturbative
part.

The Lorentz structures corresponding to
$\rho^{(GG)}_{i}\,\,(i=1,4,5,6,7,9,11,12,14,15)$ are
\begin{eqnarray}
\rho^{GG}_{\mu \nu \rho}&=&i {\rho'}^{GG}_{1} \epsilon_{\mu \nu \rho
\lambda}p_1^{\lambda}+i {\rho'}^{GG}_{4} \epsilon_{\mu \nu \rho
\lambda}p_2^{\lambda}+i {\rho'}^{GG}_{5} \epsilon_{\mu \nu \alpha
\beta}p_1^{\alpha} p_1^{\beta} {p_1}_{\nu}+i {\rho'}^{GG}_{6}
\epsilon_{\mu \rho \alpha \beta}p_1^{\alpha} p_1^{\beta}
{p_1}_{\rho} +{\rho'}^{GG}_{7}g_{\mu \nu}{p_1}_{\rho} \nonumber
\\
&&+{\rho'}^{GG}_{9}g_{\nu \rho} {p_1}_{\mu}+{\rho'}^{GG}_{11}g_{\mu
\rho} {p_2}_{\nu}+{\rho'}^{GG}_{12}g_{\nu
\rho}{p_2}_{\mu}+{\rho'}^{GG}_{14}g_{\mu \rho}
{p_2}_{\nu}+{\rho'}^{GG}_{15}g_{\nu \rho}{p_2}_{\mu}+....
\end{eqnarray}
After some long and tedious calculations, we   find that all of the
above coefficients ${\rho'}^{GG}_{i} $ are  zero. This is completely
the same as for the $\Pi_{\mu\nu}$ case.  Therefore, only the
perturbative part survives after performing the double Borel
transformation on the two variables $p_1^2$ and $p_2^2$ at the
leading order of the    heavy quark mass expansion and QCD running
coupling constant expansion for the three-point function $\Pi_{\mu
\nu \rho}$.

\section{Numerical results of form factors in QCD sum rules}
\label{Numerical results}

Now we are able to calculate form factors numerically. Firstly, we
collect the input parameters used in this work as below
\cite{ioffe 2, PDG, korner}
\begin{equation}
\begin{array}{ll}
m_c(m_c)=1.275 \pm 0.015 \rm{GeV}, & m_s(1 {\rm{GeV}})=142
{\rm{MeV}},
\\
m_u(1 {\rm{GeV}})=2.8 {\rm{MeV}}, &  m_d(1 {\rm{GeV}})=6.8
{\rm{MeV}},
\\
\alpha_{s}(1 {\rm{GeV}})=0.517, & m_{J/\psi}=3.097 \rm{GeV},
\\
m_{D^{-}}=1.869 \rm{GeV}, &  m_{D^{-}_{s}}=1.968 \rm{GeV},
\\
m_{D^{*-}}=2.010 \rm{GeV}, & m_{D^{*-}_{s}}=2.112 \rm{GeV},
\\
f_{J/\psi}=337 ^{+12}_{-13} \rm{MeV},    &
f_{D^{-}}=166^{+9}_{-10} \rm{MeV},
\\
f_{D_s^{-}}=189^{+9}_{-10}  \rm{MeV},   & f_{D^{*-}}=240^{+10}_{-10}
\rm{MeV},
\\
f_{D^{*-}_s}=262^{+9}_{-12}  \rm{MeV}.
\end{array}
\end{equation}
All the QCD parameters are adopted at the renormalization scale
around 1 GeV. It should be pointed out that the mass of charm quark
used in this work is determined form the charmonium spectrum in Ref.
\cite{ioffe 2}. As for the decay constants of charmed mesons, on the
one hand, there is a flood of papers on the theoretical
investigation of leptonic decay constants of $D^{+}$ and $D_s$
\cite{ali khan,choi,ebert,aubin,chiu,UKQCD, bordes,narison,becirevic
1,becirevic 2,rolf,khodjamirian}; on the other hand, the
measurements of decay constants of pseudoscalar $D^{+}$ and $D_s$
mesons have recently been improved  by the CLEO and BaBar
collaborations \cite{CLEO D,BABAR Ds}. Moreover, the CLEO
collaboration reported their work on the value of ratio
$f_{D_s^{+}}/f_{D^{+}}$ using the measurement of $D_s^{+}\rightarrow
l^{+} \nu$ channel and obtained $f_{D_s^{+}}=274 \pm 13 \pm 7
\mathrm{MeV}$ \cite{CLEO Ds1,CLEO Ds2}. However, the decay constants
of $D^{*+}$ and $D_s^{*}$ mesons have not been directly measured in
experiments so far. The only available results on $f_{D^{*+}}$ and
$f_{D_s^{*0}}$   from the Lattice QCD calculations \cite{aubin,
becirevic 2, k.c. bowler} determine  $f_{D_s^{*}}= 272 \pm 16
^{+3}_{-20} \mathrm{MeV}$ that is smaller than the value of decay
constant for $D_s^{+}$ measured by the CLEO collaboration \cite{CLEO
Ds1,CLEO Ds2}.   To reduce the theoretical uncertainties in the
three-point sum rules of the weak transition form factors, due to
quarks masses, threshold parameters and Coulomb-like corrections of
$J/\psi$ effectively \cite{kiselev Bc},  we use the decay constants
$f_{\psi}$ and $f_{D^{(*)-}_{d,s}}$ calculated from the two-point
QCD sum rules in leading order of $\alpha_s$, the same as that in
the  three-point sum rules.  The explicit calculations of the decay
constants, in the framework of QCD sum rules,  for both $J/\psi$ and
$D^{(\ast)}_{d,s}$ are displayed in Appendix \ref{Decay constants of
charmonium and charmed meson}. Our results indicae that
${f_{D_s^{\ast}} \over f_{D^{\ast}}}\simeq {f_{D_s} \over
f_{D}}=1.1$, which are in good agreement with that from lattice
simulation \cite{k.c. bowler} and experiments \cite {CLEO Ds1,CLEO
Ds2}.

For the threshold parameters $s_1^0$ and $s_2^0$, one should
determine them by demanding the QCD sum rules results to be
relatively stable in allowed regions for $M_1^2$ and $M_2^2$, the
values of which should be around the mass square of the
corresponding first excited states. As for the heavy-light mesons,
the standard value of the threshold in the $X$ channel would be
$s^0_{X}=(m_X+\Delta_X)^2$, where $\Delta_X$ is about $0.6$ GeV
\cite{dosch, matheus, bracco, navarra, Colangelo}, and we simply
take it as $(0.6 \pm 0.1)\; \mathrm{GeV}$ for the error estimate in
the numerical analysis. When it comes to the heavy quarkonium,
following the method in Ref. \cite{matheus, bracco, Colangelo}, we
select the effective threshold parameter to ensure the appearance of
the pleasant platform and also around the mass square of $\psi(2S)$.
In this way,  the contributions from both the excited states
including $\psi(2S)$ and the continuum states are contained in the
spectral function.

\subsection{The numerical results of the form factors}

\subsubsection{Evaluation of the form factors for the $J/\psi \to D^{-}$}

With   all the parameters listed above, we can obtain the numerical
values of the form factors. The form factors should not depend on
the Borel masses $M_1$ and $M_2$ in a complete theory. However, as
we truncate the operator product expansion up to dimension-5 and
keep the perturbative expansion in $\alpha_s$ to leading order, an
obvious dependence of the form factors on these two Borel parameters
would emerge. Therefore, one should look for a region where the
results only mildly vary with respect to the Borel masses, so that
the truncation is reasonable and acceptable.

 With a careful analysis, $s_1^0=13.7$
GeV$^2$ and  $s_2^0=6.1$ GeV$^2$  are chosen  for the form factor
$V$ calculation. We require the contributions from the higher states
to be less than 30 \% and the value of $V$  does not vary
drastically within the selected region for the Borel masses. As
commonly understood, the Borel parameters $M_1^2$ and $M_2^2$ should
not be too large in order to insure that the contributions from the
higher excited states and continuum are not too significant. On the
other hand, the Borel masses also could not be too small for the
sake of validity of OPE in the deep Euclidean region, since the
contributions of higher dimension operators pertain to the higher
orders in ${1 \over M_i}(i=1,2) $. Different from that adopted in
previous literature \cite{p. ball,weak decays of QCDSR 1} where the
ratio of $M_1$ and $M_2$ was fixed, in the calculation of  form
factors, we let $M_1$ and $M_2$ vary independently as suggested by
the authors of Ref. \cite{yangkc,kiselev}. In this way, we indeed
find a Borel platform $M_1^2 \in [6.0, 10.0] \mathrm{GeV}^2, M_2^2
\in [1.0, 2.0] \mathrm{GeV}^2$,   plotted in Fig.~\ref{all form
factor of J psi to D},  which satisfy the conditions discussed
above. One can directly read from this figure that $V(q^2=0)$ is
$0.48^{+0.07}_{-0.05}$,   whose uncertainties originate from the
variation of the Borel parameters.

\begin{figure}[tb]
\begin{center}
\begin{tabular}{ccc}
\includegraphics[scale=0.5]{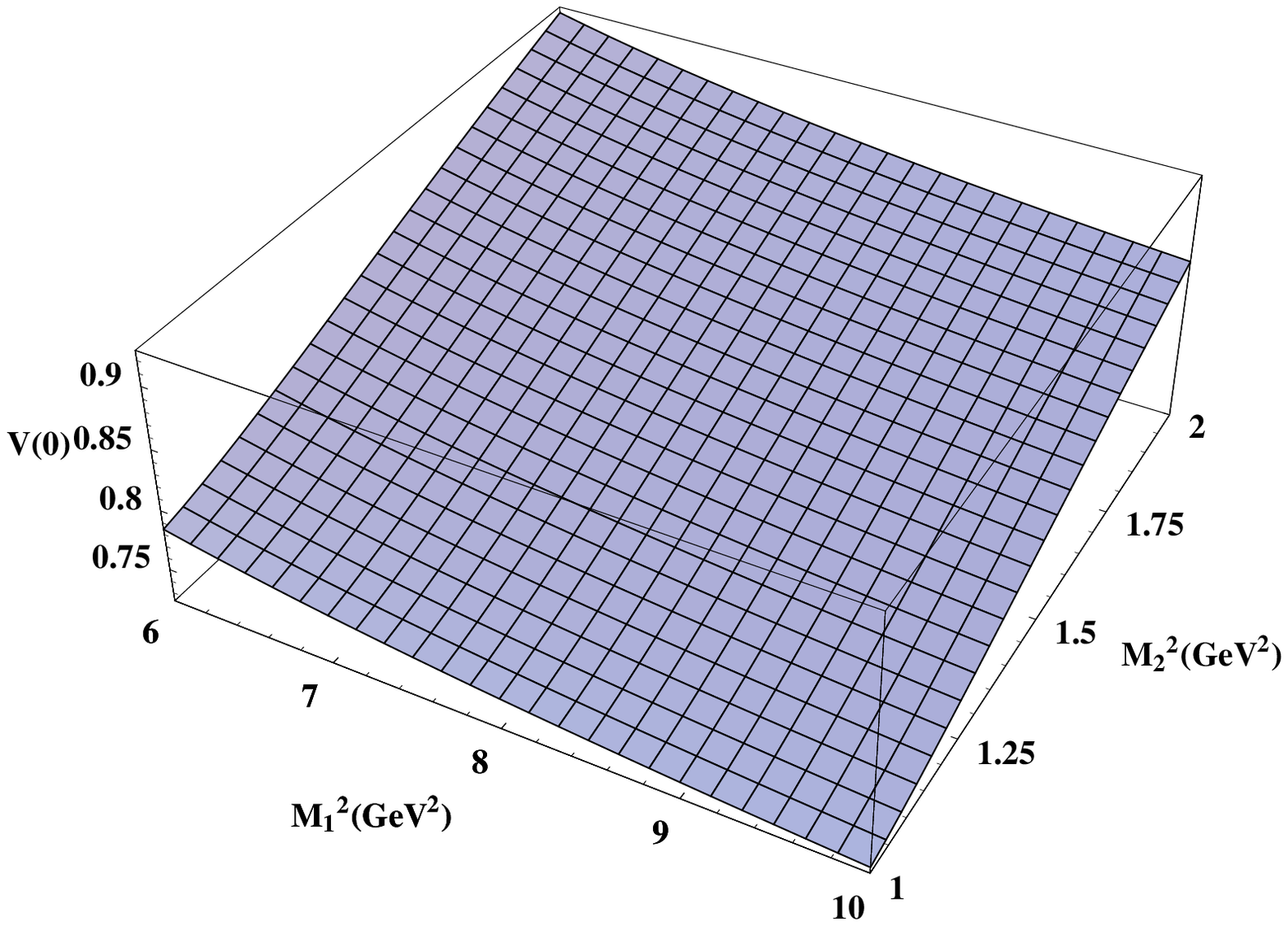}
\includegraphics[scale=0.5]{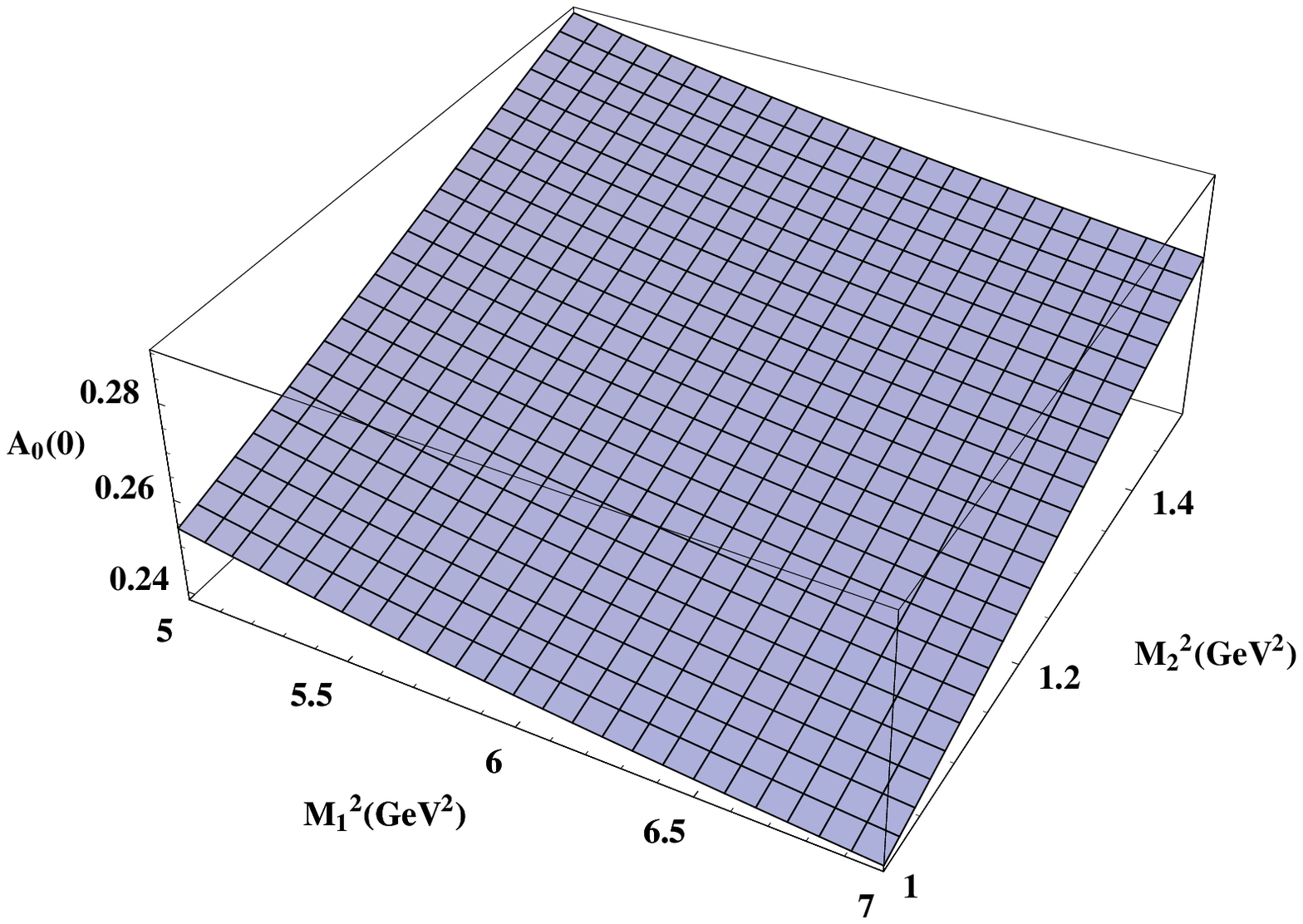}
\\
\includegraphics[scale=0.5]{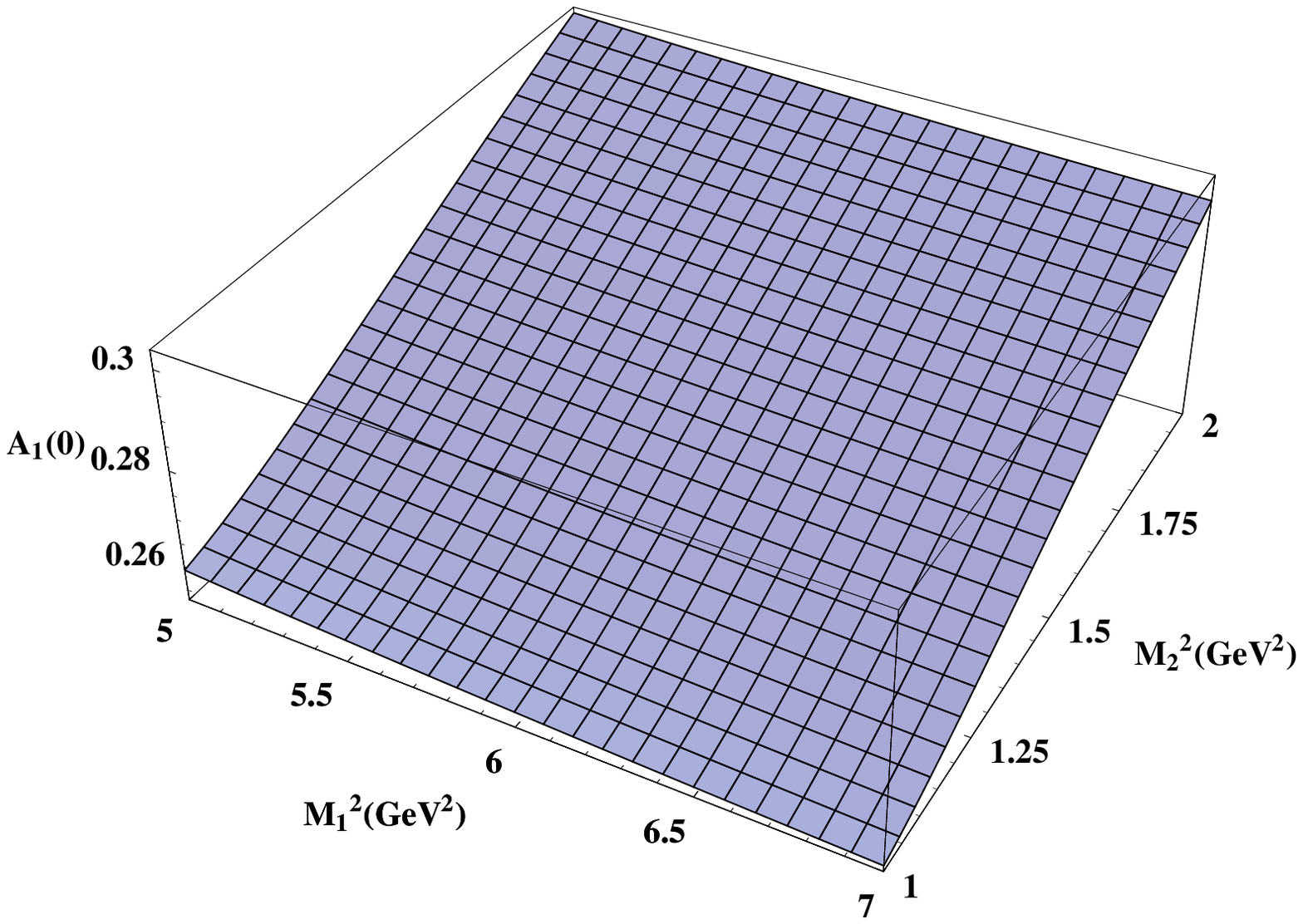}
\includegraphics[scale=0.5]{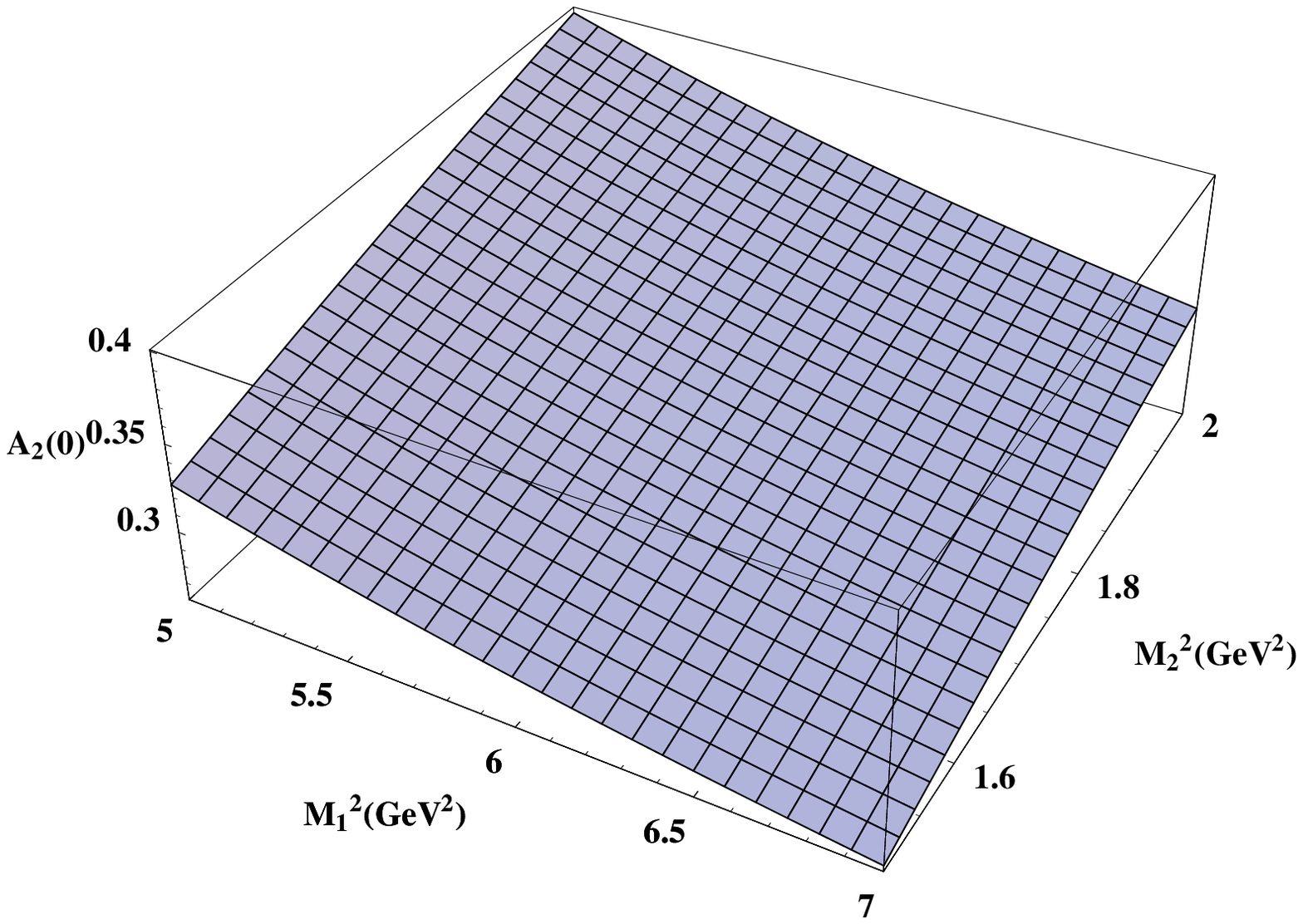}
\end{tabular}
\caption{Dependence of form factors $V, A_0, A_1$ and $A_2$ at
$q^2=0$ responsible for the decay of $J/\psi \to D^{-} $ on the
Borel masses. }\label{all form factor of J psi to D}
\end{center}
\end{figure}

Following the same procedure, we also obtain numerical results for
the other three form factors $A_0,\;A_1$ and $A_2$ within the chosen
Borel window as  shown in Fig. \ref{all form factor of J psi to D}.
The numerical results of form factors $V, A_0, A_1$ and $A_2$ at
zero momentum transfer are then
\begin{eqnarray}
V(0)=0.81^{+0.12}_{-0.08},  \,\,\, A_0(0)=0.27^{+0.02}_{-0.03},
\,\,\, A_1(0)=0.27^{+0.03}_{-0.02},
\,\,\,A_2(0)=0.34^{+0.07}_{-0.07} . \label{44}
\end{eqnarray}
 It needs to be emphasized that the form factors $A_1(q^2)$,
$A_2(q^2)$ and $A_0(q^2)$ should satisfy the relation
$(m_{\psi}+m_{D^{-}})A_1(0)+(m_{\psi}-m_{D^{-}})A_2(0)=2
m_{\psi}A_0(0)$ to ensure disappearance of the divergence at the
pole $q^2=0$. The theoretical uncertainties in the form factors
 (\ref{44})  originate from the Borel masses
$M_1^2$ and $M_2^2$. They are at the level of 15\%, which implies
stable results from the  QCD sum rules approach.

Indeed there are some extra errors originating from the values of
$s_1^0$ and $s_2^0$ which correspond to the threshold of the higher
excited resonances and continuum states for the $J/\psi$ and $D$
channels respectively. In the  QCD sum rules approach, the values of
threshold parameter is usually in the vicinity of mass square of the
first physical excited state, therefore, we do not investigate the
dependence of form factors on the threshold parameter in this work
as that in Ref. \cite{Coulomb corrections,kiselev}, where a larger
threshold value of charmonium is adopted. This uncertainty would
cause errors in the resultant form factors. Besides, the
fluctuations of the charm quark mass can also result in the
uncertainties of the form factors, which is evaluated to be at the
level of $6-8\%$. Moreover, the input parameters such as the decay
constants of $D$ meson and $J/\psi$ can also bring on additional
uncertainties. Combing the errors from various parameters discussed
above, the uncertainties on the form factors can be estimated within
20 to 30\%, expected by the general understanding of the theoretical
framework.

Next, we can further investigate the $q^2$ dependence of the form
factors $V, A_0,A_1$ and $A_2$. The physical region of $q^2$ for
$J/\psi \to D^{-} l^+ \nu_l$ is $0 \leq q^2
\leq(m_{J/\psi}-m_{D^{-}})^2 \simeq 1.5 \mathrm{GeV}^2$. However,
with the QCD sum rules, we could not obtain the form factors in the
whole physical region, since the additional singularities - so
called ``non-Landau-type" singularities emerge, which had been
extensively discussed in Ref.\cite{weak decays of QCDSR 1}. To avoid
this kind of singularity, we restrict our calculations in the range
of $q^2 \in [0, 0.47] \rm{GeV}^2$. We show the $q^2$ dependence of
the form factors $V, A_0,A_1$ and $A_2$ in Fig. \ref{t dependence of
J psi to D}.

\begin{figure}[tb]
\begin{center}
\begin{tabular}{ccc}
\includegraphics[scale=0.6]{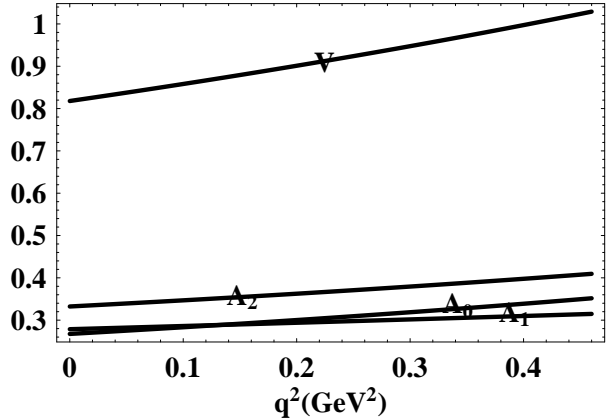}
\end{tabular}
\caption{$q^2$ dependence of form factors $V, A_0, A_1$ and $A_2$
for $J/\psi \to D^{-}$ within the kinematical region without
non-Landau-type singularities.}\label{t dependence of J psi to D}
\end{center}
\end{figure}

In addition, for the convenience of applications to phenomenology,
one can parameterize the above form factors in a three-parameter
form \cite{khodjamirian heavy falvors}:
\begin{eqnarray}\label{form}
F_{i}(q^2)={F_i(0) \over 1-a_i q^2/m_{D^{-}}^{2}+b_i
q^4/m_{D^{-}}^{4}},
\end{eqnarray}
where $F_i$ denotes the form factors $V, A_0, A_1$ and $A_2$, and
$a_i$ and $b_i$ are the parameters to be fixed. Using the QCD sum
rules $F_i(q^2)$ with $q^2$   restricted within a certain kinematic
region,  we can fix the parameters $a_i$, $b_i$ in the expression.
This double-pole expression for form factors can be generalized to
the whole kinematic region. Finally, our results for the parameters
$a_i$, $b_i$ are given as
\begin{eqnarray}
a_{V}=1.65^{+0.20}_{-0.03}, \qquad b_{V}=0.76^{+0.44}_{-0.09},
\qquad
a_{A_0}=1.97^{+0.15}_{-0.03}, \qquad b_{A_0}=1.19^{+0.31}_{-0.05},  \nonumber \\
a_{A_1}=0.93^{+0.27}_{-0.12}, \qquad b_{A_1}=0.46^{+0.29}_{-0.01},
\qquad a_{A_2}=1.47^{+0.14}_{-0.16}, \qquad
b_{A_2}=0.32^{+0.19}_{-0.21}.
\end{eqnarray}
For the other form factors which are discussed in the following
subsections, we will adopt the same procedure to obtain the form
factors in the whole kinematic region.

\subsubsection{$J/\psi \to D_{s}^-$ form factors}

\begin{figure}[tb]
\begin{center}
\begin{tabular}{ccc}
\includegraphics[scale=0.7]{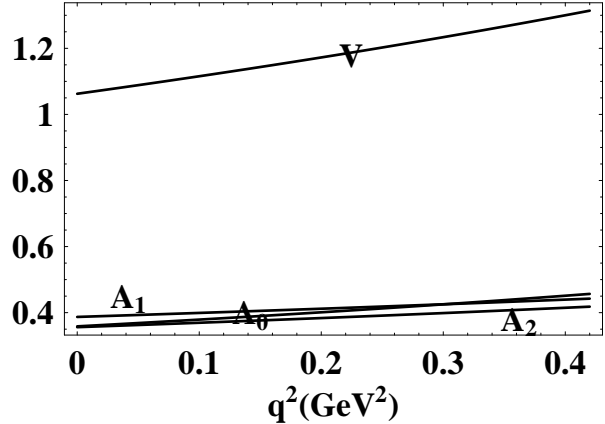}
\end{tabular}
\caption{$q^2$ dependence of form factors $V, A_0, A_1$ and $A_2$
for $J/\psi \to D^{-}_{s} $ within the kinematical region without
non-Landau-type singularities.}\label{t dependence of J psi to Ds}
\end{center}
\end{figure}

Now, we move on to the computations of the form factors for the
transition  $J/\psi \to D^{-}_{s}$, which is quite similar to that
for $J/\psi \to D^{-}$, only with $d$ quark in $D^{-}$ being
replaced by  $s$. It is also noted that the threshold parameter
$s_2^0=6.6 \rm{GeV}^2$ for the $D_s$ channel and the Borel window
are shifted slightly compared with that of $J/\psi \to D^{-}$. Since
the figures are very similar to the case for $J/\psi \to D^{-}$, we
just omit them. The obtained form factors for $J/\psi \to D_{s}^-$
at $q^2=0$ are
\begin{eqnarray}
V(0)=1.07^{+0.05}_{-0.02},  \,\,\, A_0(0)=0.37^{+0.02}_{-0.02},
\,\,\, A_1(0)=0.38^{+0.02}_{-0.01},
\,\,\,A_2(0)=0.35^{+0.08}_{-0.07}.
\end{eqnarray}
And the parameters $a_i$ and $b_i$ defined above for the $q^2$
dependence formula (with the replacement of $D \to D_s$) are fixed
as
\begin{eqnarray}
a_{V}=1.86^{+0.26}_{-0.03}, \qquad b_{V}=0.90^{+0.43}_{-0.04},  \qquad
a_{A_0}=2.12^{+0.0}_{-0.04}, \qquad b_{A_0}=1.30^{+0.0}_{-0.04},  \nonumber \\
a_{A_1}=1.18^{+0.24}_{-0.01}, \qquad b_{A_1}=0.27^{+0.29}_{-0.04},
\qquad a_{A_2}=1.41^{+0.20}_{-0.29}, \qquad
b_{A_2}=0.38^{+0.15}_{-0.01}.
\end{eqnarray}
Again, the relation
$(m_{\psi}+m_{D_s})A_1(0)+(m_{\psi}-m_{D_s})A_2(0)=2 m_{\psi}A_0(0)$
is well respected, which guarantees that the hadronic matrix element
responsible for the $J/\psi \to D^{-}_{s} $ is free of divergence
due to the pole at $q^2=0$. We show the dependence of the form
factors on $q^2$ in Fig.\ref{t dependence of J psi to Ds}.

\subsubsection{$J/\psi \to D^{*-}$ form factors}

The evaluation of the form factors responsible for $J/\psi \to
D^{*-}$ is performed following the standard procedure,  with
appropriate Borel windows   obtained. The threshold value for
$D^{*-}$ channel takes $s_2^0=6.8 \rm{GeV}^2$ in our numerical
analysis. The form factors at   zero momentum transfer $q^2=0$ are
collected below as
\begin{eqnarray}
 \tilde{A}_1(0)&=&0.40^{+0.03}_{-0.01}, \qquad
 \tilde{A}_2(0)=0.44^{+0.10}_{-0.04},  \qquad
 \tilde{A}_3(0)=0.86^{+0.05}_{-0.01}, \qquad
 \tilde{A}_4(0)=0.91^{+0.06}_{-0.04}, \nonumber \\
 \tilde{V}_1(0)&=&0.41^{+0.01}_{-0.01},  \qquad
 \tilde{V}_2(0)=0.63^{+0.01}_{-0.04}, \qquad
 \tilde{V}_3(0)=0.22^{+0.03}_{-0.01},\qquad
 ~\tilde{V}_4(0)=0.26^{+0.03}_{-0.05}, \nonumber \\
 \tilde{V}_5(0)&=&1.37^{+0.08}_{-0.03}, \qquad
 \tilde{V}_6(0)=0.87^{+0.05}_{-0.01}.
\end{eqnarray}
 From the above results, we find that the form factors obtained in
the QCD sum rules respect the relations
$\tilde{A}_1(0)=\tilde{A}_2(0)$ and $\tilde{V}_3(0)=\tilde{V}_4(0)$,
which are essential to assure that the hadronic matrix element of
$J/\psi \to D^{*-}$ is free of divergence at $q^2=0$.

Different from that discussed for the $J/\psi \to D^{-}_{d,s}$ case,
not all the form factors which appear in the hadronic matrix element
for $J/\psi \to D^{*-}$, are suitably parameterized in the form of
eq.(\ref{form}) with the thee-parameter approximation. To be more
specific, the $q^2$ dependence of the form factors
$\tilde{A}_1(q^2)$ and $\tilde{A}_2(q^2)$ are written in the
following form \cite{burford,h.y. cheng}
\begin{eqnarray}\label{av1}
 F_{i}(q^2)={F_i(0) \over (1-a_i
q^2/m_{D^{*-}}^{2})^2},
\end{eqnarray}
where $F_i$ represents $\tilde{A}_1$ and $\tilde{A}_2$; while the
other eight form factors are written in the three-parameter form,
\begin{eqnarray}\label{a2}
G_{i}(q^2)={G_i(0) \over 1-a_i q^2/m_{D^{*-}}^{2}+b_i
q^4/m_{D^{*-}}^{4}},
\end{eqnarray}
where $G_i$ can be $\tilde{A}_3$, $\tilde{A}_4$ and
$\tilde{V}_i(i=1-6)$. We then extend the form factors to the whole
physical region $0 \leq q^2 \leq(m_{J/\psi}-m_{D^{*-}})^2 \simeq 1.2
\mathrm{GeV}^2$, by fitting the parameters as
\begin{eqnarray}
 &&a_{\tilde{A}_1}=1.77^{+0.04}_{-0.01}, \qquad
 a_{\tilde{A}_2}=1.95^{+0.17}_{-0.25}, \qquad
 a_{\tilde{A}_3}=2.93^{+0.18}_{-0.08}, \qquad
 b_{\tilde{A}_3}=2.47^{+0.54}_{-0.27}, \nonumber \\
 &&a_{\tilde{A}_4}=2.78^{+0.05}_{-0.03}, \qquad
 b_{\tilde{A}_4}=1.78^{+0.27}_{-0.14}, \qquad
 ~a_{\tilde{V}_1}=1.96^{+0.03}_{-0.03}, \qquad
 b_{\tilde{V}_1}=0.98^{+0.07}_{-0.06}, \nonumber \\
 &&a_{\tilde{V}_2}=2.11^{+0.04}_{-0.04}, \qquad
 b_{\tilde{V}_2}=0.21^{+0.05}_{-0.02}, \qquad
 ~a_{\tilde{V}_3}=1.92^{+0.0}_{-0.03},  \qquad
 b_{\tilde{V}_3}=1.87^{+0.12}_{-0.12}, \nonumber \\
 &&a_{\tilde{V}_4}=2.96^{+0.34}_{-0.23}, \qquad
 b_{\tilde{V}_4}=1.97^{+1.03}_{-0.34}, \qquad
 ~a_{\tilde{V}_5}=1.92^{+0.0}_{-0.12}, \qquad
 b_{\tilde{V}_5}=1.03^{+0.0}_{-0.21}, \nonumber \\
 &&a_{\tilde{V}_6}=2.00^{+0.27}_{-0.02}, \qquad
 b_{\tilde{V}_6}=1.08^{+0.53}_{-0.03}
 .
\end{eqnarray}

\begin{figure}[tb]
\begin{center}
\begin{tabular}{ccc}
\includegraphics[scale=0.6]{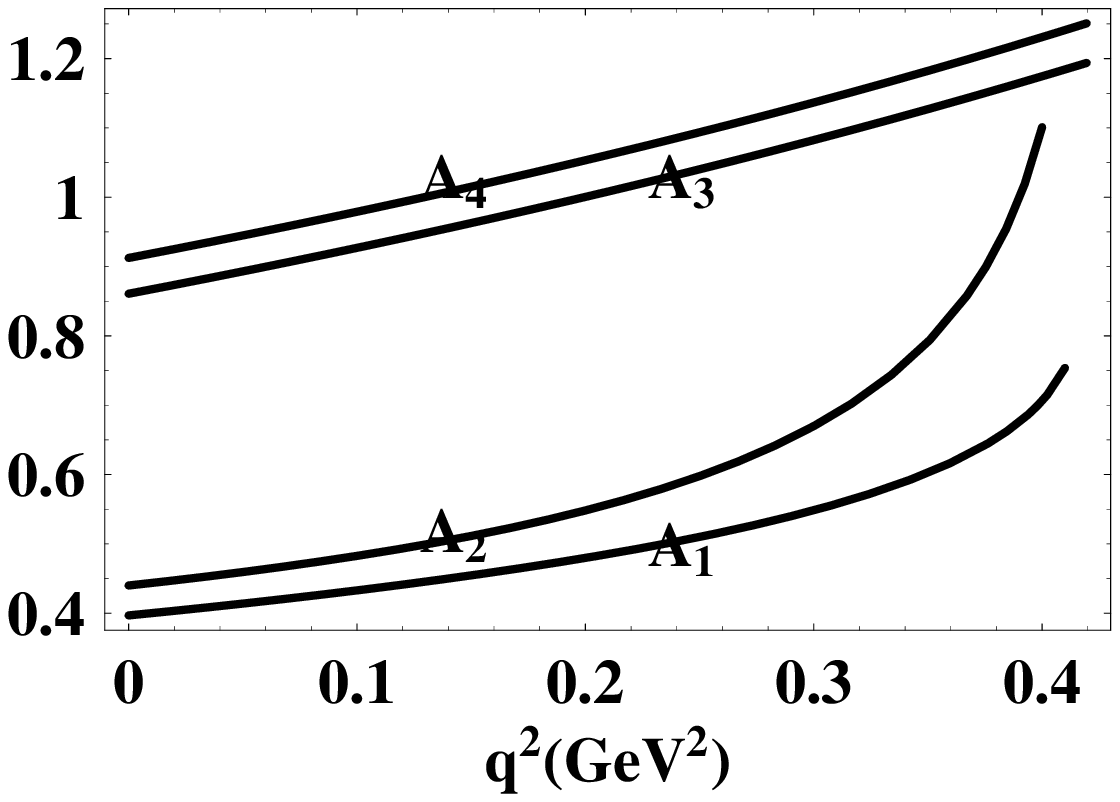}
\includegraphics[scale=0.6]{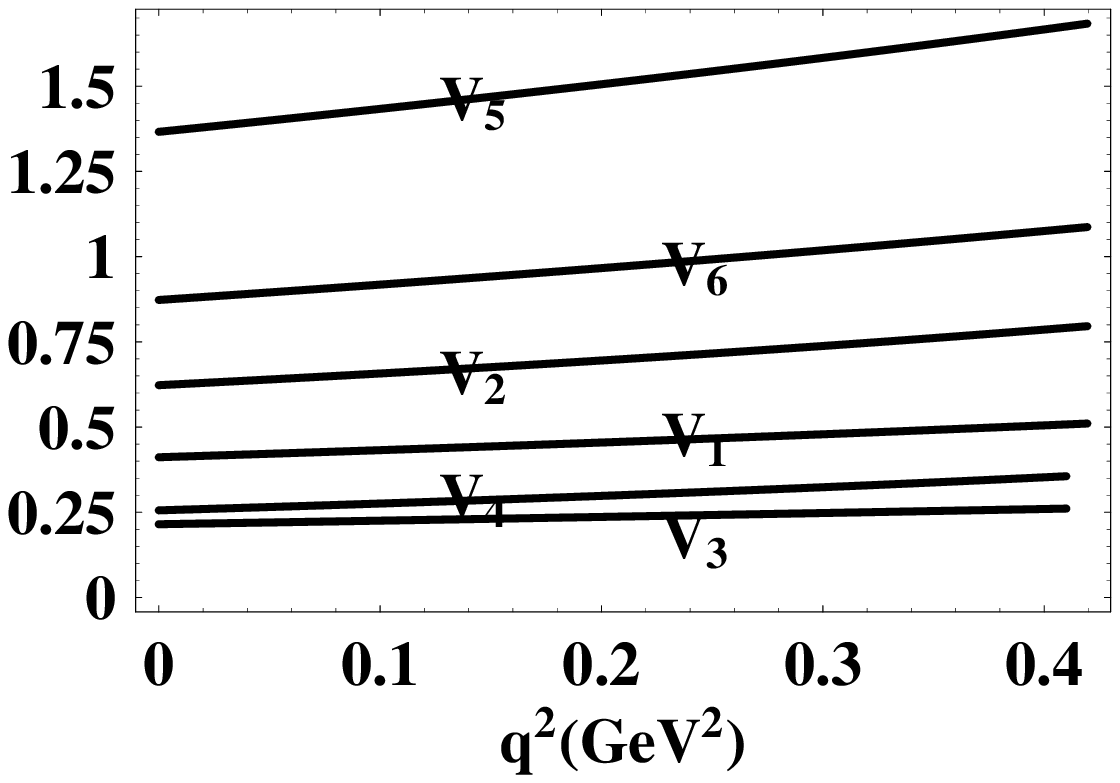}
\end{tabular}
\caption{$q^2$ dependence of form factors $\tilde{A}_1, \tilde{A}_2,
\tilde{A}_3, \tilde{A}_4 ,\tilde{V}_1, \tilde{V}_2,
\tilde{V}_3,\tilde{V}_4, \tilde{V}_5$ and $\tilde{V}_6$ for $J/\psi
\to D^{*-}$ within the kinematical region without non-Landau-type
singularities.}\label{t dependence of J psi to Dstar}
\end{center}
\end{figure}

\subsubsection{The form factors for $J/\psi \to D^{*-}_{s}$ }

The computation on the amplitude of $J/\psi \to D^{*-}_{s} $ is
almost the same as that for  $J/\psi \to D^{*-}$, only $d$ quark in
$D^{*-}$ being replaced by $s$ quark, with the difference resulting
in a different Borel platform. Besides, the threshold parameter for
the $D^{*-}_{s}$ channel is set as $s_2^0= 7.4 \rm{GeV}^2$ in the
calculations. The $q^2$ dependence of the form factors falling into
the region of $q^2 \in [0, 0.37] \rm{GeV}^2$ is plotted in Fig.
\ref{t dependence of J psi to Ds star}. As mentioned before, the
form factors $\tilde{A}_1$ and $\tilde{A}_2$ can be parameterized in
the form of Eq.(\ref{av1}), while the other form factors can be fit
in the usual three-parameter form in Eq.(\ref{a2}). The parameters
$a_i$ and $b_i$ can be determined by reproducing the numbers
obtained from the QCD sum rules for the kinematic region $q^2 \in
[0, 0.37] \rm{GeV}^2$ and then we generalize the results to the
whole physical region $q^2 \in [0, 0.97] \rm{GeV}^2$. The values of
these parameters together with the form factors at $q^2=0$ are
collected
 for convenience as,

\begin{eqnarray}
 &&a_{\tilde{A}_1}=1.92^{+0.0}_{-0.05},  \qquad
 a_{\tilde{A}_2}=1.85^{+0.01}_{-0.21}, \qquad
 a_{\tilde{A}_3}=3.07^{+0.12}_{-0.01}, \qquad
 b_{\tilde{A}_3}=1.98^{+0.80}_{-0.16}, \nonumber \\
 &&a_{\tilde{A}_4}=3.08^{+0.06}_{-0.02}, \qquad
 b_{\tilde{A}_4}=2.08^{+0.60}_{-0.26}, \qquad
 a_{\tilde{V}_1}=2.05^{+0.13}_{-0.02}, \qquad
 b_{\tilde{V}_1}=0.90^{+0.29}_{-0.06}, \nonumber \\
 &&a_{\tilde{V}_2}=2.53^{+0.06}_{-0.12}, \qquad
 b_{\tilde{V}_2}=0.07^{+0.17}_{-0.35}, \qquad
 ~a_{\tilde{V}_3}=2.04^{+0.16}_{-0.12}, \qquad
 b_{\tilde{V}_3}=2.14^{+0.07}_{-0.0},  \nonumber \\
 &&a_{\tilde{V}_4}=3.32^{+0.37}_{-0.33}, \qquad
 b_{\tilde{V}_4}=1.76^{+1.58}_{-0.67}, \qquad
 ~a_{\tilde{V}_5}=1.92^{+0.22}_{-0.03}, \qquad
 b_{\tilde{V}_5}=0.81^{+0.40}_{-0.05}, \nonumber \\
 &&a_{\tilde{V}_6}=2.00^{+0.26}_{-0.04}, \qquad
 b_{\tilde{V}_6}=0.81^{+0.55}_{-0.15},
\end{eqnarray}
and
\begin{eqnarray}
\tilde{A}_1(0)&=&0.53^{+0.03}_{-0.01},  \qquad
\tilde{A}_2(0)=0.53^{+0.05}_{-0.01}, \qquad
\tilde{A}_3(0)=0.91^{+0.05}_{-0.01}, \qquad
\tilde{A}_4(0)=0.91^{+0.06}_{-0.01},
\nonumber \\
\tilde{V}_1(0)&=&0.54^{+0.01}_{-0.01},  \qquad
\tilde{V}_2(0)=0.69^{+0.05}_{-0.06}, \qquad
\tilde{V}_3(0)=0.24^{+0.03}_{-0.01},\qquad
~\tilde{V}_4(0)=0.26^{+0.03}_{-0.03},
\nonumber \\
\tilde{V}_5(0)&=&1.69^{+0.10}_{-0.03}, \qquad
\tilde{V}_6(0)=1.14^{+0.08}_{-0.01}.
\end{eqnarray}
In the same way, the relations $\tilde{A}_1(0)=\tilde{A}_2(0)$ and
$\tilde{V}_3(0)=\tilde{V}_4(0)$ are well satisfied.

\begin{figure}[tb]
\begin{center}
\begin{tabular}{ccc}
\includegraphics[scale=0.6]{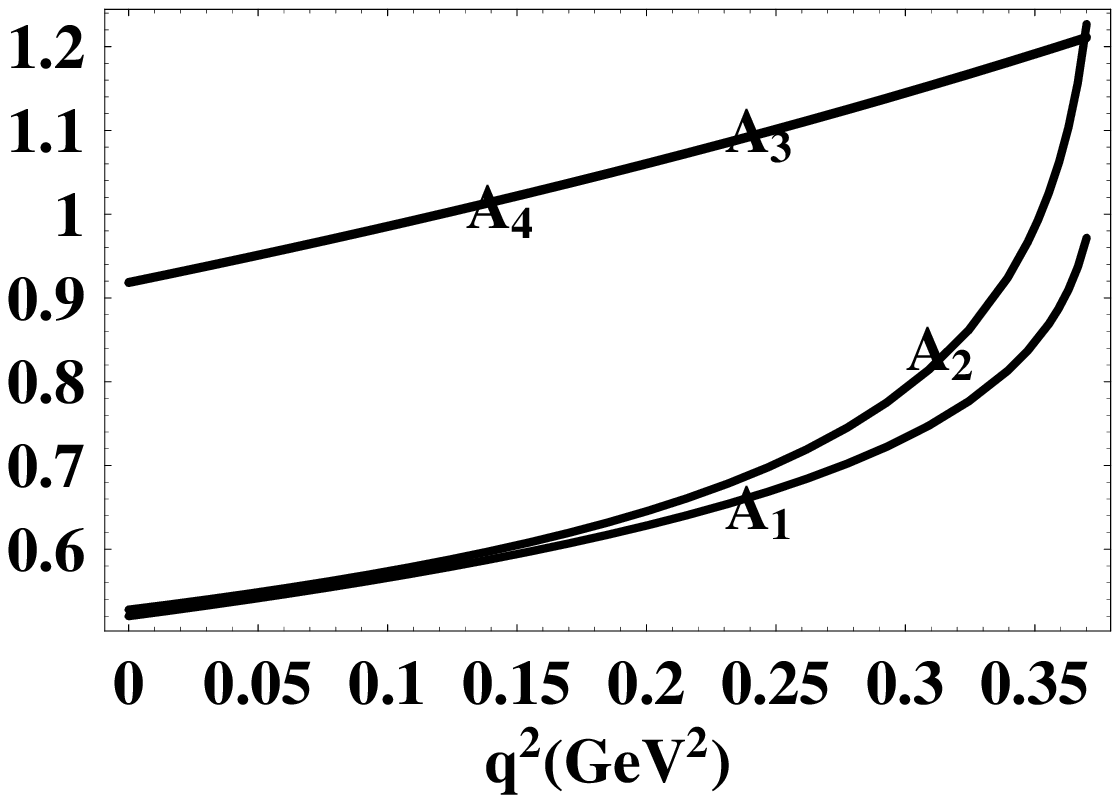}
\includegraphics[scale=0.6]{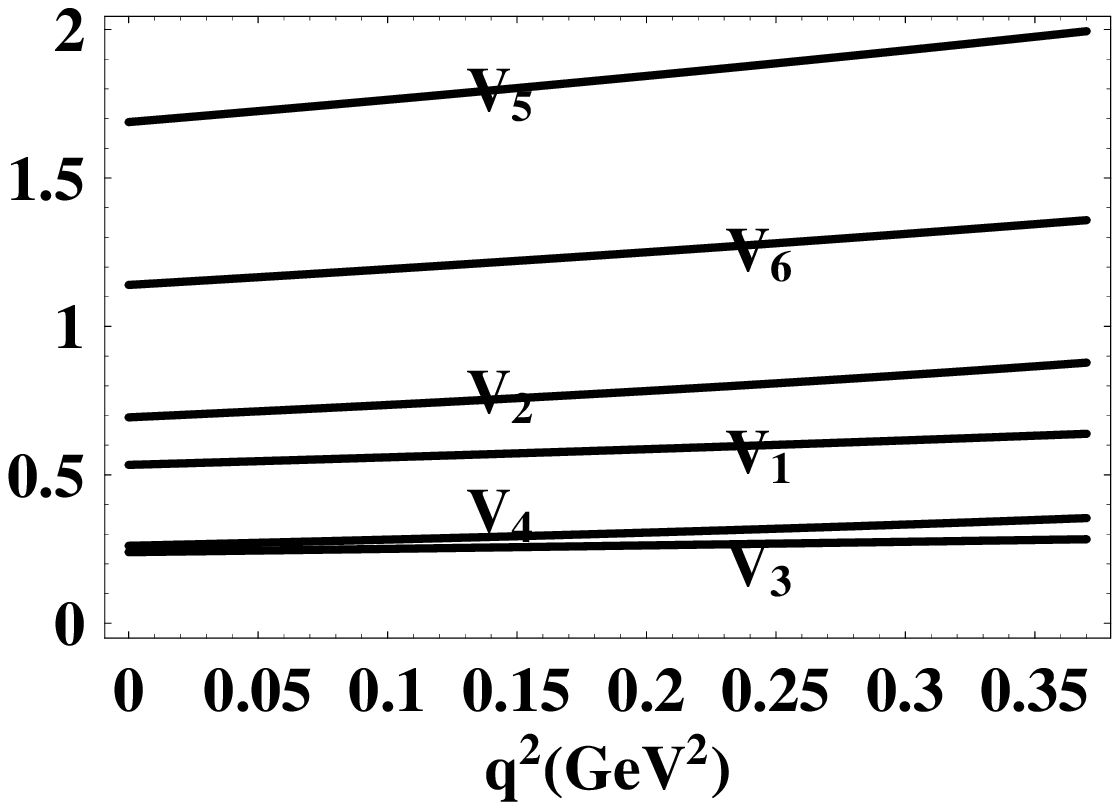}
\end{tabular}
\caption{$q^2$ dependence of form factors $\tilde{A}_1, \tilde{A}_2,
\tilde{A}_3, \tilde{A}_4 ,\tilde{V}_1, \tilde{V}_2,
\tilde{V}_3,\tilde{V}_4, \tilde{V}_5$ and $\tilde{V}_6$ for $J/\psi
\to D^{*-}_{s}$ within the kinematical region without
non-Landau-type singularities.} \label{t dependence of J psi to Ds
star}
\end{center}
\end{figure}

\section{Decay rates for semi-leptonic weak decays of $J/\psi$}
\label{decay rate}

With the form factors  derived above, we can perform calculations on
partial widths of the semi-leptonic decays of $J/\psi$. The relevant
CKM parameters are directly taken from the particle data book
\cite{PDG}:
\begin{equation}
\begin{array}{ll}
|V_{cd}|=0.2271, \qquad |V_{cs}|=0.973.
\end{array}
\end{equation}

For the semi-leptonic decays $J/\psi \to D^{(*)-}_{(d,s)} l^{+}
\nu_{l}$ ($l= e, \mu$), the differential partial decay rate is
written as
\begin{equation}
{d\Gamma_{\psi \to D^{(*)-}_{(d,s)} l^{+} \nu_{l}} \over d q^2} ={1
\over 3}{1 \over (2 \pi)^3} {1 \over 32 m_{\psi}^3}
\int_{u_{min}}^{u_{max}} |{\widetilde{M}}_{\psi \to D^{(*)-}_{(d,s)}
l^{+} \nu_{l}}|^2 du,
\end{equation}
where $u=(p_{l^+}+p_{\nu_l})^2$;  $p_{l^+}$ and $p_{\nu_l}$ are the
momenta of $l^{+}$ and $\nu_{l}$ respectively; $|\widetilde{M}|^2$
is square of the transition amplitude after integrating over the
angle between the $l^{+}$ and $D^{(*)-}_{(d,s)}$. The upper and
lower bounds for $u$ are given as
\begin{equation}
u_{min}= {m_{\psi}^2 -m_{D^{(*)}_{(d,s)}}^2 -q^2 \over 2
\sqrt{q^2}},\qquad u_{max}={q^2+m_l^2 \over 2\sqrt{q^2}}.
\end{equation}
The transition amplitude $\widetilde{M}$ for $J/\psi \to
D^{(*)-}_{(d,s)} l^{+} \nu_{l}$ reads as
\begin{eqnarray}
\widetilde{M}_{\psi \to D^{(*)-}_{(d,s)} l^{+} \nu_{l}}={G_{F} \over
\sqrt{2}} V_{cq}^{*} \langle
D^{(*)}_{(d,s)}|\bar{q}\gamma_{\mu}(1-\gamma_5)c|J/\psi\rangle
\bar{\nu}_l \gamma^{\mu}(1-\gamma_5)l.
\end{eqnarray}
Finally we get the branching ratios of the semi-leptonic decays as
\begin{equation}
\begin{array}{ll}
{\rm{BR}}(J/\psi \to D^{-} e^{+} \nu_{e})=7.3 ^{+4.3}_{-2.2} \times
10^{-12}, & {\rm{BR}}(J/\psi \to D^{-} {\mu}^{+}
\nu_{\mu})=7.1^{+4.2}_{-2.2} \times 10^{-12}\!,
\nonumber \\
{\rm{BR}}(J/\psi \to D^{-}_{s} e^{+} \nu_{e})=1.8^{+0.7}_{-0.5}
\times 10^{-10}, & {\rm{BR}}(J/\psi \to D^{-}_{s} {\mu}^{+}
\nu_{\mu})=1.7^{+0.7}_{-0.5} \times 10^{-10},
\nonumber \\
{\rm{BR}}(J/\psi \to D^{*-} e^{+} \nu_{e})=3.7 ^{+1.6}_{-1.1}\times
10^{-11}, & {\rm{BR}}(J/\psi \to D^{*-} {\mu}^{+}
\nu_{\mu})=3.6^{+1.6}_{-1.1} \times 10^{-11},
\nonumber \\
{\rm{BR}}(J/\psi \to D^{*-}_{s} e^{+} \nu_{e})=5.6^{+1.6}_{-1.6}
\times 10^{-10}, & {\rm{BR}}(J/\psi \to D^{*-}_{s} {\mu}^{+}
\nu_{\mu})=5.4^{+1.6}_{-1.5} \times 10^{-10},
\end{array}
\end{equation}
where we have combined various uncertainties in the form factors
discussed in last section  to determine the final error tolerance in
our theoretical calculations. Our predictions are much below the
present experimental upper bounds \cite{BES}: ${\rm{BR}}(J/\psi \to
D^{-}_{s} e^{+} \nu_{e}+c.c.)<4.9\times 10^{-5}$, ${\rm{BR}}(J/\psi
\to D^{-} e^{+} \nu_{e}+c.c.)<1.2\times 10^{-5}$.

A few remarks are presented in order. Firstly,  the sum of the
branching fractions of semi-leptonic decays of $J/\psi$ whose final
state includes $D^{-}_s$, $D^{*-}_s$, $e$ and $\mu$ and their charge
conjugate channels can reach as large as $3.1 \times 10^{-9}$, which
is expected to be marginally observed at BESIII. Secondly, it is
worthwhile to point out that the decay rates for the dominant
semi-leptonic weak decays of $J/\psi$ obtained in Ref.
\cite{Sanchis-Lonzano} were about $7 \times 10^{-9}$, which are two
times greater than that calculated in this work. This discrepancy
can attribute to the heavy quark spin symmetry and the non-recoil
approximation used in Ref. \cite{Sanchis-Lonzano}, also to the
different methods used to estimate the non-perturbative form factors
\footnote {In Ref.\cite{Sanchis-Lonzano}, the ISGW model \cite{ISGW}
was employed to compute the single form factor $\eta_{12}$, while we
adopt the QCD sum rules to calculate the form factors in this
work.}. Thirdly, the ratio of $R_1 \equiv {{\rm{BR}}(J/\psi \to
D^{*-}_{s} e^{+} \nu_{e}) \over {\rm{BR}}(J/\psi \to D^{-}_{s} e^{+}
\nu_{e})}\simeq 3.1$ is about 2 times larger than the value
calculated in \cite{Sanchis-Lonzano}, where the assumption of heavy
quark spin symmetry and non-recoil approximation were adopted.
Forth, the ratios $R_2 \equiv {{\rm{BR}}(J/\psi \to D^{-}_{s} e^{+}
\nu_{e}) \over {\rm{BR}}(J/\psi \to D^{-} e^{+} \nu_{e})}$ and $R_3
\equiv {{\rm{BR}}(J/\psi \to D^{*-}_{s} e^{+} \nu_{e}) \over
{\rm{BR}}(J/\psi \to D^{*-} e^{+} \nu_{e})}$ should be equal to
$|{V_{cs} \over V_{cd}}|^2 \simeq 18.4$ under the SU(3) flavor
symmetry limit. Our numerical calculations show that $R_2  \simeq
24.7$ and $R_3  \simeq 15.1 $,  which implies large effect of SU(3)
symmetry breaking.

\section{Discussions and conclusions}
\label{Discussions and conclusions}

The charmonium $J/\psi$ meson can decay  via the strong and
electromagnetic interactions, thus weak decays of $J/\psi$ should be
very rare unless there is new physics beyond the standard model to
make a substantial contribution. If such weak decays can be measured
by the future experiments with sizable branching ratios, it would be
a clear signal for new physics.

To make the new physics signal clearly  distinguishable from the
standard model, a careful study of weak decays in SM is needed. In
this work, we calculated the form factors of weak transitions of
$J/\psi\to D^{(*)}_{(d,s)}$ in terms of the QCD sum rules. With the
form factors, we estimate the branching ratios of the semileptonic
weak decays of $J/\psi$ and find that the sum of the branching
ratios corresponding to the dominant modes is about $3.1 \times
10^{-9}$ which   may be marginally measured by BES III. The  QCD sum
rules approach possesses uncontrollable errors   as large as 20 to
30\%,   was confirmed by our numerical results. Moreover, due to a
Coulomb-type correction in heavy quarkonium decay (or $B_c$), which
may manifest as the ladder structure in the loop-triangle (as a part
of multi-loop diagrams), the spectral function needs to be
multiplied by a finite renormalization factor \cite{QCDSR 1,kiselev,
Coulomb corrections, Kiselev prd}. This would bring up another kind
of uncertainty. It is expected that this kind of correction can give
birth to the twice multiplication of the form factors at the maximal
momentum transfer. However, in this work, we calculate both the
three-point QCD sum rules of the weak transition form factors and
the two-point sum rules for the decay constant of $J/\psi$ to the
same order of $\alpha_s$. Then, it is expected that most
uncertainties due to the Coulomb-like corrections are canceled in
our calculations; therefore, the Coulomb-like corrections for the
$J/\psi$ channel are not included in our calculations. As for the
heavy-light mesons, there are no corrections in the power of inverse
velocity for it, since the light quark moves relativistically.
Therefore, one should explore the sum rules for both three-point and
two-point correlation functions up to next-to-leading order in
strong coupling constant so that Coulomb-like corrections to the
heavy-light mesons can be canceled effectively. Moreover, an
explicit calculation of Coulomb-like corrections to the heavy-light
vertex in triangle diagram is still not available now, which can be
left for further considerations.


One can trust the numerical results to a certain accuracy, at least
the order of magnitude is reliable. With these form factors we may
continue to estimate the rates of non-leptonic weak decays of
$J/\psi$ as long as the factorization theorem is proved. That would
be the contents of our next work \cite{Wangetal}.

The branching ratios of semi-leptonic weak decays of $J/\psi$  are
very small in SM, even though their strong decay modes are
OZI-suppressed. Our numerical results indicate that even with a
large database which will be collected by the BES III, such weak
decay modes may only be marginally observed.
Therefore we lay our expectation on our BES III colleagues
and hope them to provide sufficiently large database to make this
challengeable field more fruitful.

\section*{Acknowledgements}

This work is partly supported by National Science Foundation of
China under Grant No.10475085, 10625525 and  10475042. The authors
would like to thank T.M. Aliev, P. Colangelo, T. Huang, V.V.
Kiselev, N. Paver, M.Z. Yang, F.K. Guo, Y.L. Shen and  W. Wang for
helpful discussions.

\appendix

\section{The explicit forms of  Wilson coefficients for $\Pi_{\mu \nu}$}
\label{Wilson coefficients for D}

In this appendix, we would like to show the explicit expressions of
Wilson coefficients appearing in  Eq.(\ref{V pseudoscalar}-\ref{A0
pseudoscalar}) after the Borel transformation. As mentioned before,
only perturbative part contributes to the correlation function for
$J/\psi$ decays to $D^{-}_{d,s}$ at the leading order of heavy quark
mass expansion and the $\alpha_s$ expansion series. Namely, Eq.
(\ref{fi expansion}) can be written as
\begin{eqnarray}
f_i=f_i^{pert} {\mathbf{I}}+ O(\alpha_s)+ O({1 / m_{h}}),
\end{eqnarray}
with $h$ being the heavy charm quark here. The $f_i^{pert}$ can be
related to $\rho_{i}^{pert}$ defined in Eq. ({\ref{spectral density
1}}) by
\begin{eqnarray}
f_i^{pert}=\int^{s_2^0}_{(m_c+m_q)^2} ds_2  \int^{s_1^0}_{s_1^L}
ds_1 {\rho_{i}^{pert}(s_1,s_2,q^2) \over (s_1-p_1^2) (s_2-p_2^2)},
\end{eqnarray}
or
\begin{eqnarray}
\hat{\mathcal{B}}f_i^{pert}= \int^{s_2^0}_{(m_c+m_q)^2} ds_2
\int^{s_1^0}_{s_1^L} ds_1 {1 \over M_1^2}   {\mathrm{e}}^{-s1/M_1^2}
{1 \over M_2^2} {\mathrm{e}}^{-s_2/M_2^2}
{\rho_{i}}^{pert}(s_1,s_2,q^2).
\end{eqnarray}
The lowest bound of $s_1$, i.e., $s_1^L$ can be determined by the
Eq. (\ref{integral region}) as
\begin{eqnarray}
s_1^L&=&- {1 \over 2 m_q^2}\bigg[m_c^4 -(2 m_q^2 +s_2 +q^2)m_c^2+m_q^2+s_2
q^2-m_q^2(s_2+q^2) \nonumber \\
&&+\sqrt{m_c^4-2(m_q^2+s_2)m_c^2+(m_q^2-s_2)^2}
\sqrt{m_c^4-2(m_q^2+q^2)m_c^2+(m_q^2-q^2)^2} \bigg], \label{s1L}
\end{eqnarray}
according to the Landau equation \cite{landau equation, S matrix}.

The obvious forms of $\rho_{i}^{pert}\,\, (i=0,2,4,5)$ are
\begin{eqnarray}
\rho_{0}^{pert}(s_1,s_2,q^2)&=&-{3 \over 4 \pi^2 \lambda^{3/2}} [m_c
\lambda + (m_c-m_q) s_1 (2 m_c^2-2 m_q^2 -s_1+s_2+q^2)],\nonumber
\\
\rho_{2}^{pert}(s_1,s_2,q^2)&=&- {3 s_1 \over 2 \pi^2 \lambda^{5/2}}
\bigg\{ (m_c-m_q)s_1 \bigg[6 m_c^4- 6(2 m_q^2 +s_1-s_2)m_c^2 +6
m_q^4
+(s_1-s_2)^2  \nonumber \\
&&+6 m_q^2(s_1-s_2)\bigg]+m_c [2(m_c-m_q)(2m_c+m_q)-s_1+s_2]\lambda
\nonumber \\
&&+q^2[2(m_c-m_q)s_1(3m_c^2 -3 m_q^2 -s_1 +2 s_2) +m_c \lambda
+(m_c-m_q)s_1 q^2]\bigg\},\nonumber
\\
\rho_{4}^{pert}(s_1,s_2,q^2)&=&{ 3 \over 4 \pi^2 \lambda^{5/2}}
\bigg\{[-m_c \lambda^2+(m_c-m_q)(2s_2 m_c^2 +s_1(2 m_q^2 +s_1-s_2)]
\lambda\nonumber \\
&& +2 (m_c-m_q) s_1 [3(s_1+s_2)m_c^4 \nonumber \\
&&-2 (3(s_1+s_2)m_q^2+(s_1-s_2)(s_1+2 s_2))m_c^2+(s_1-s_2)^2 s_2
+3m_q^4 (s_1+s_2)\nonumber  \\
&& +2 m_q^2 (s_1-s_2)(s_1+2 s_2)] +(m_c-m_q)q^2 [2s_1 (s_2^2 +(-2 m_c^2 +2 m_q^2+s_1)s_2
\nonumber  \\
&& +(m_c^2-m_q^2)(-3 m_c^2+3m_q^2 +4 s_1))-(2 m_c^2+s_1)\lambda  -4
s_1 (m_c^2 -m_q^2 +s_2)q^2]\bigg\},\nonumber
\\
\rho_{5}^{pert}(s_1,s_2,q^2)&=&{3 \over 8 \pi^2 \lambda^{3/2}}
\bigg\{\lambda (m_q s_1 +m_c s_2 -m_c q^2) - 2 (m_c-m_q)[\lambda
m_c^2
\nonumber  \\
&&+(m_c^2-m_q^2)s_1 (m_c^2-m_q^2 -s_1+s_2)+s_1
(m_c^2-m_q^2+s_2)q^2]\bigg\}.
\end{eqnarray}
In this appendix, we adopt the notion $\lambda \equiv
\lambda(s_1,s_2,q^2)$ for the convenience of writing.

\section{The expressions of Wilson coefficients for $\Pi_{\mu \nu \rho}$}
\label{Wilson coefficients for Dstar}

Similarly, we will display the   forms of Wilson coefficients
emerged in the Eq.(\ref{A1 vector}-\ref{V6 vector}) after performing
the Borel transformation. As been discussed in the text, only
perturbative part contributes to the three-point function,
\begin{eqnarray}
F_i=F_i^{pert} {\mathbf{I}}+ O(\alpha_s)+ O({1 / m_{h}}).
\end{eqnarray}
The relationship between $F_i^{pert}$, and ${\rho'}_i^{pert}$ are
given as
\begin{eqnarray}
F_i^{pert}&=&\int^{s_2^0}_{(m_c+m_q)^2} ds_2  \int^{s_1^0}_{s_1^L}
ds_1 {{\rho'}_{i}^{pert}(s_1,s_2,q^2) \over (s_1-p_1^2) (s_2-p_2^2)}
\end{eqnarray}
or
\begin{eqnarray}
\hat{\mathcal{B}}F_i^{pert}&=& \int^{s_2^0}_{(m_c+m_q)^2} ds_2
\int^{s_1^0}_{s_1^L} ds_1 {1 \over M_1^2}   {\mathrm{e}}^{-s1/M_1^2}
{1 \over M_2^2} {\mathrm{e}}^{-s_2/M_2^2}
{{\rho'}_{i}}^{pert}(s_1,s_2,q^2)
\end{eqnarray}
where the definition of $s_1^{L}$ is given in Eq.(\ref{s1L}). The
manifest expressions of
${\rho'}_i^{pert}(i=1,4,5,6,7,9,11,12,14,15)$ are displayed as
\begin{eqnarray}
{\rho'}_{1}^{pert}(s_1,s_2,q^2)&=&\frac{3}{4\pi^2\lambda^{5/2}}
\{-4s_1^2m_c^4+(8m_q^2s_1^2+(s_1-s_2)(s_1+s_2)^2-(s_1-3s_2)\lambda)m_c^2\nonumber\\
&&+2m_q(s_1+s_2)\lambda
m_c-4m_q^4s_1^2+s_2(3s_1+s_2)(\lambda-(s_1-s_2)^2)-m_q^2(s_1+s_2)(s_1^2-s_2^2+\lambda)
\nonumber\\
&&+q^2[4s_1m_c^4+(3s_2^2+10s_1s_2-s_1(8m_q^2+s_1)+\lambda)m_c^2-2m_q\lambda
m_c+3s_2(s_1+s_2)^2\nonumber\\
&&+4m_q^4s_1-s_2\lambda+m_q^2(s_1^2-10s_1s_2-3s_2^2+\lambda)+(m_c^2-m_q^2+s_2)
q^2(-s_1-3s_2+q^2)]\},\nonumber
\\
{\rho'}_{4}^{pert}(s_1,s_2,q^2)&=&\frac{3s_1}{4\pi^2\lambda^{5/2}}
\{-4s_2m_c^4-2(-4s_2m_q^2+s_1^2+3s_2^2-4s_1s_2-\lambda)m_c^2+4m_qm_c\lambda-4m_q^4s_2
\nonumber\\
&&-(s_1+3s_2)((s_1-s_2)^2-\lambda)+2m_q^2
(s_1^2-4s_1s_2+3s_2^2-\lambda)+q^2(4m_c^4+4(s_2-2m_q^2)m_c^2\nonumber\\
&&+4m_q^4+3(s_1+s_2)^2
-4m_q^2s_2-\lambda+q^2(2m_c^2-2m_q^2-3s_1-s_2+q^2))\},\nonumber
\\
{\rho'}_{5}^{pert}(s_1,s_2,q^2)&=&\frac{1}{4\pi^2\lambda^{5/2}}
\{-(s_1+s_2-q^2)[3(m_c^2-m_q^2+s_2)q^4-6((2s_1+s_2)m_c^2+s_2(s_1+s_2)\nonumber\\
&&-m_q^2
(2s_1+s_2))q^2-3(2s_1m_c^4-(4s_1m_q^2+3s_1^2+s_2^2-\lambda)m_c^2\nonumber\\
&&+2m_qm_c\lambda+2m_q^4s_1
+m_q^2(3s_1^2+s_2^2-\lambda)+s_2(s_1^2-s_2^2+\lambda))]-2s_2[3s_1(-2(m_c^2-m_q^2)^2
\nonumber\\
&&+s_1^2-s_2^2-4(m_c^2-m_q^2)s_2)-3(2m_c^2+s_1)\lambda+3s_1
q^2(q^2-2(s_1+s_2))]\},\nonumber
\\
{\rho'}_{6}^{pert}(s_1,s_2,q^2)&=&\frac{3}{4\pi^2\lambda^{5/2}}
\bigg\{-2s_1(3s_1+s_2)m_c^4+2\bigg[s_1(2m_q^2+s_1-s_2)(3s_1+s_2)-(2s_1+s_2)
\lambda\bigg]m_c^2\nonumber\\
&&-4m_qs_1m_c\lambda
+s_1\bigg[-2(3s_1+s_2)m_q^4+2(-3s_1^2+2s_1s_2+s_2^2+\lambda)m_q^2+
(s_1-s_2)^2(s_1+s_2)\nonumber\\
&&-(s_1+3s_2)\lambda\bigg]+q^2\bigg[2s_1m_c^4+2(\lambda-2s_1(m_q^2+2s_1))m_c^2\nonumber\\
&&+s_1(2m_q^4+8s_1m_q^2-(s_1+s_2)(3s_1+5s_2)+\lambda)
+s_1(2m_c^2-2m_q^2+3s_1+5s_2-q^2)q^2\bigg]\bigg\},\nonumber
\\
{\rho'}_{7}^{pert}(s_1,s_2,q^2)&=&-\frac{3}{16\pi^2\lambda^{5/2}}
\bigg\{(2(m_c-m_q)^2-2s_2)\lambda^2
-2\bigg[s_2(-s_1+s_2-q^2)+(m_c^2-m_q^2)(s_1+s_2-q^2)\bigg]\nonumber\\
&&\times(s_1-s_2+q^2)\lambda-8\bigg[(m_c^2-m_q^2+s_2)(s_1+s_2-q^2)-2s_1s_2\bigg]
\bigg[((s_1+s_2-q^2)^2\nonumber\\
&&-4s_1s_2)m_c^2+s_1(m_c^2-m_q^2+s_2)^2+s_1^2s_2-s_1(m_c^2-m_q^2+s_2)
(s_1+s_2-q^2)\bigg]\bigg\},\nonumber
\\
{\rho'}_{9}^{pert}(s_1,s_2,q^2)&=&-\frac{3}{8\pi^2\lambda^{5/2}}
\bigg\{-((m_c-m_q)^2-s_2)\lambda^2-\bigg[s_2(-s_1+s_2-q^2)+(m_c^2-m_q^2)
(s_1+s_2-q^2)\bigg]\nonumber\\
&&\times(s_1-s_2+q^2)\lambda+2\bigg[\lambda
m_c^2+(m_c^2-m_q^2)s_1(m_c^2-m_q^2-s_1+s_2)+s_1(m_c^2-m_q^2+s_2)q^2\bigg]\lambda\nonumber\\
&&-4\bigg[(m_c^2-m_q^2+s_2)(s_1+s_2-q^2)-2s_1s_2\bigg]\bigg[((s_1+s_2-q^2)^2-4s_1s_2)m_c^2
\nonumber\\
&&+s_1(m_c^2-m_q^2+s_2)^2+s_1^2s_2-s_1(m_c^2-m_q^2+s_2)
(s_1+s_2-q^2)\bigg]\bigg\},\nonumber
\\
{\rho'}_{11}^{pert}(s_1,s_2,q^2)&=&-\frac{3s_1}{8\pi^2\lambda^{5/2}}
\bigg\{-\lambda^2-(-2m_c^2+2m_q^2+s_1-s_2-q^2)[-2(m_c-m_q)^2
-s_1+s_2+q^2]\lambda \nonumber\\
&&-4\bigg[((s_1+s_2-q^2)^2-4s_1s_2)m_c^2+s_1(m_c^2-m_q^2+s_2)^2+s_1^2s_2\nonumber\\
&&-s_1(m_c^2-m_q^2+s_2)(s_1+s_2-q^2)\bigg](-2m_c^2+2m_q^2+s_1-s_2-q^2)\bigg\},\nonumber
\\
{\rho'}_{12}^{pert}(s_1,s_2,q^2)&=&-\frac{3}{8\pi^2\lambda^{5/2}}
\bigg\{s_1\lambda^2-s_1(-2m_c^2+2m_q^2+s_1-s_2-q^2)(s_1+s_2-q^2)\lambda\nonumber\\
&&+2\bigg[\lambda
m_c^2+(m_c^2-m_q^2)s_1(m_c^2-m_q^2-s_1+s_2)+s_1(m_c^2-m_q^2+s_2)q^2\bigg]\lambda
\nonumber\\
&&-4s_1\bigg[((s_1+s_2-q^2)^2-4s_1s_2)m_c^2+s_1(m_c^2-m_q^2+s_2)^2+s_1^2s_2
\nonumber\\
&&-s_1(m_c^2-m_q^2+s_2)(s_1+s_2-q^2)\bigg]
(-2m_c^2+2m_q^2+s_1-s_2-q^2)\bigg\},\nonumber
\\
{\rho'}_{14}^{pert}(s_1,s_2,q^2)&=&-\frac{3}{4\pi^2\lambda^{7/2}}
\bigg\{-2m_c^2(s_1+s_2-q^2)\lambda^2-s_1\bigg[4s_1s_2(-2m_c^2+2m_q^2+s_1-s_2-q^2)\nonumber\\
&&-(s_2(-s_1+s_2-q^2)+(m_c^2-m_q^2)(s_1+s_2-q^2))(s_1+s_2-q^2)\nonumber\\
&&+3(m_c^2-m_q^2+s_2)(s_1+s_2-q^2)(2m_c^2-2m_q^2-s_1+s_2+q^2)\bigg]\lambda
\nonumber\\
&&+4s_1\bigg[4s_2\{(s_1+s_2-q^2)^2+s_1s_2\}s_1^2-3(m_c^2-m_q^2+s_2)\{(s_1+s_2-q^2)^2
\nonumber\\
&&+6s_1s_2\}(s_1+s_2-q^2)s_1+2(m_c^2(s_1+s_2-q^2)^4
+\{6s_1(m_c^2-m_q^2+s_2)^2
\nonumber\\
&&-2m_c^2s_1s_2\}(s_1+s_2-q^2)^2
+2s_1^2s_2(3(m_c^2-m_q^2+s_2)^2-4m_c^2s_2))-2(m_c^2-m_q^2\nonumber\\
&&+s_2)\{3((s_1+s_2-q^2)^2-4s_1s_2)m_c^2+5s_1(m_c^2-m_q^2+s_2)^2\}
(s_1+s_2-q^2)\bigg]\bigg\},\nonumber
\\
{\rho'}_{15}^{pert}(s_1,s_2,q^2)&=&-\frac{3}{4\pi^2\lambda^{7/2}}
\bigg\{-2m_c^2(s_1+s_2-q^2)\lambda^2-s_1(4s_1s_2(-2m_c^2+2m_q^2+s_1-s_2-q^2)\nonumber\\
&&-\bigg[s_2(-s_1+s_2-q^2)+(m_c^2-m_q^2)(s_1+s_2-q^2)\bigg](s_1+s_2-q^2)\nonumber\\
&&+3(m_c^2-m_q^2+s_2)(s_1+s_2-q^2)(2m_c^2-2m_q^2-s_1+s_2+q^2))\lambda\nonumber\\
&&+4\bigg[-10s_2^2(s_1+s_2-q^2)s_1^3+12s_2(m_c^2-m_q^2+s_2)((s_1+s_2-q^2)^2+s_1s_2)s_1^2
\nonumber\\
&&-3(2s_2((s_1+s_2-q^2)^2-4s_1s_2)m_c^2+(m_c^2-m_q^2+s_2)^2((s_1+s_2-q^2)^2+6s_1s_2))
(s_1\nonumber\\
&&+s_2-q^2)s_1+2(m_c^2-m_q^2+s_2)(m_c^2(s_1+s_2-q^2)^4+2s_1((m_c^2-m_q^2+s_2)^2\nonumber\\
&&-m_c^2s_2)
(s_1+s_2-q^2)^2+2s_1^2s_2((m_c^2-m_q^2+s_2)^2-4m_c^2s_2))\bigg]\bigg\}.
\end{eqnarray}
\label{perturbative 2}

\section{Decay constants of $J/\psi$ and  $D^{(\ast)}_{d,s}$ in two-point QCD sum rules}
\label{Decay constants of charmonium and charmed meson}

In this appendix, we would like to collect the sum rules for the
decay constants of $J/\psi$ and  $D^{(\ast)}_{d,s}$ for the
completeness of the paper. The decay constant of $J/\psi$  in the
two-point QCD sum rules can be written as \cite{reinders,matheus}

\begin{eqnarray}
f_{\psi}^2 m_{\psi}^2 e^{- {m_{\psi}^2 \over M^2}}
=\int_{4m_c^2}^{s^0_{\psi}} ds {s \over 4 \pi^2}\ (1-{4 m_c^2 \over
s})^{1/2} (1+{2 m_c^2 \over s})e^{-{s \over M^2}} +[-{m_c^2 \over 4
M^2} +{1 \over 16} +{1 \over 48}e^{-{4 m_c^2 \over M^2}}] \langle 0
| {\alpha_s \over \pi} G_{\mu \nu}^2|0 \rangle,
\end{eqnarray}
where the nonrelativistic approximation for the gluon condensate has
been adopted for the convenience of performing the Borel
transformation. It is observed that the gluon condensate has tiny
effect on the results of the form factors and hence are neglected in
the sum rules of charmed mesons. The non-perturbative condensates
used in the evaluation of the sum rues can be grouped as
\begin{equation}
\begin{array}{ll}
\langle 0|\bar{q} q|0\rangle=-(1.65 \pm 0.15) \times 10^{-2}
{\rm{GeV}}^3 (q=u, d), & \langle 0|\bar{s} s|0\rangle=(0.8 \pm 0.1)
\langle 0|\bar{q} q|0\rangle,
\\
\langle 0|{\alpha_s \over \pi}G_{\mu \nu}^2|0\rangle=0.005 \pm 0.004
\rm{GeV}^2, & \langle 0|\bar{q}_i \sigma \cdot G q_i|0\rangle=m_0^2
\langle 0|\bar{q}_i q_i|0\rangle,
\end{array}
\end{equation}
where $m_0^2=(0.8 \pm 0.2) {\rm{GeV}}^2$ and the subscript $``i"$
denotes the flavor of quarks. Based on the two-point sum rules of
$J/\psi$ and the parameters showed above, we can derive the decay
constant of $J/\psi$ as $337 ^{+12}_{-13} {\rm{MeV}}$, where we have
combined the uncertainties from the variations of the Borel masses
and threshold value for $J/\psi$ channel.

The sum rules for the decay constant of $D_{q}$ can be given by
\cite{reinders, reinders P, belyaev, bediaga}
\begin{eqnarray}
{m_{D_q}^4 \over (m_c +m_q) ^2} f_{D_q}^2 e^{-{m_{D_q}^2 \over M^2}}
&=&{3 \over 8 \pi^2} \int_{(m_c +m_q)^2}^{s^0_{D_{q}}} ds [1- {
(m_c-m_q)^2 \over s}] \lambda^{1/2}(s,m_c^2,m_q^2) e^{-{s \over
M^2}}  \\
&&+\left(-m_c +{m_q \over 2}+{m_q m_c^2 \over 2M^2}\right)
e^{-{m_c^2 \over M^2}} \langle 0 | \bar{q} q  |0 \rangle -{m_c \over
2M^2} (1- {m_c^2 \over 2M^2})  e^{-{m_c^2 \over M^2}} \langle
0|\bar{q} \sigma \cdot G q|0\rangle, \nonumber
\end{eqnarray}
 from which we can arrive at the decay constants of pseudoscalar
charmed mesons as $f_{D_d}= 166 ^{+9}_{-10} {\rm{MeV}}$ and
$f_{D_s}= 189 ^{+9}_{-10} {\rm{MeV}}$.

The decay constants of vector charmed mesons $f_{D_{q}^{\ast}}$ in
the framework of QCD sum rules can be calculated as \cite{reinders,
reinders V, belyaev}
\begin{eqnarray}
f_{D_{q}^{\ast}}^2 m_{D_{q}^{\ast}}^2 e^{-{ m_{{D_{q}^{\ast}}}^2
\over M^2}} &=& {1 \over 8 \pi^2}
\int_{(m_c+m_q)^2}^{s^0_{{D_{q}^{\ast}}}} ds
\lambda^{1/2}(s,m_c^2,m_q^2) \left[2- {m_c^2+m_q^2-6 m_c m_q \over
s} -{(m_c^2-m_q^2)^2\over s^2}\right] e^{-{s \over
M^2}}  \nonumber \\
&&+ \left\{\left[-(m_c +{8 \over 3} m_q) +{1 \over 2}{m_q m_c^2
\over M^2}\right] e^{-{m_c^2 \over M^2}} +{ 2m_q (4 M^2-m_c^2)\over
3 M^2}
\right\} \langle 0|\bar{q} q|0\rangle \nonumber \\
&&+{m_c^3 \over 4 M^4} e^{-{m_c^2 \over M^2}} \langle 0|\bar{q}
\sigma \cdot G q|0\rangle,
\end{eqnarray}
 from which we can achieve the decay constants of vector charmed
mesons as $f_{D_d^{\ast}}= 240 ^{+10}_{-10} {\rm{MeV}}$ and
$f_{D_s^{\ast}}= 262 ^{+9}_{-12} {\rm{MeV}}$.


\begin{thebibliography}{99}

\bibitem{Sanchis-Lonzano}M.A. Sanchis-Lonzano, Z. Phys. {\bf C 62},
 271 (1994).

\bibitem{BES}M. Ablikim  {\it et al.}[BES Collaboration],
Phys. Lett. {\bf B 639}, 418 (2006) [arXiv: hep-ex/0604005].
\bibitem{BESIII} F.A. Harris, arXiv: hep-ex/0606059.


\bibitem{quark model}M. Wirbel, B. Stech, M. Bauer,  Z. Phys.  {\bf C 29}, 637 (1985).

\bibitem{light front} M. Terent'ev, Sov. J. Nucl. Phys. {\bf 24}, 106 (1976);
 V. Berestetsky and M. Terent'ev, $ibid$. {\bf 24}, 547 (1976); {\bf 25}, 347 (1977);
 P. Chung, F. Coester, and W. Polyzou, Phys. Lett. {\bf B  205}, 545
 (1988); W. Jaus, Phys. Rev.  {\bf D  41}, 3394 (1990); {\bf 44}, 2851 (1991);
 {\bf 60}, 054026 (1999). C. Ji, P. Chung and S. Cotanch, Phys. Rev. {\bf D 45}, 4214
 (1992); H.Y. Cheng, C.Y. Cheung and C.W. Hwang, Phys.
  Rev. {\bf D  55}, 1559 (1997) [arXiv: hep-ph/9607332];  H.Y. Cheng, C.K. Chua and C.W. Hwang, Phys.
  Rev. {\bf D 69}, 074025 (2004) [arXiv: hep-ph/0310359]; C.W. Hwang and Z.T. Wei, J. Phys.
  {\bf G  34}, 687 (2007) [arXiv: hep-ph/0609036]; C.D. L\"{u}, W. Wang and Z.T. Wei,
Phys. Rev.  {\bf D  76}, 014013 (2007) [arXiv: hep-ph/0701265].

\bibitem{QCDSR 1}M. A. Shifman, A. I. Vainshtein and V. I. Zakharov,
Nucl. Phys. {\bf B 147}, 385 (1979).


\bibitem{QCDSR 2}V. A. Novikov, M. A. Shifman, A. I. Vainshtein and V. I. Zakharov,
Nucl.Phys. {\bf B 191}, 301 (1981).

\bibitem{PQCD}
 Y.Y. Keum, H.N. Li, and A.I. Sanda, Phys. Lett. {\bf B 504}, 6 (2001)
  [arXiv: hep-ph/0004004];
 C.D. L\"{u}, K. Ukai, and M.Z. Yang, Phys. Rev. {\bf D 63}, 074009 (2001) [arXiv:
  hep-ph/0004213];
 T. Kurimoto, H.n. Li, A.I. Sanda, Phys. Rev. {\bf D 65}, 014007 (2002) [arXiv: hep-ph/0105003];
Phys. Rev. {\bf D 67}, 054028 (2003) [arXiv: hep-ph/0210289];  Z.T.
Wei, M.Z. Yang, Nucl. Phys. {\bf B 642}, 263 (2002) [arXiv:
hep-ph/0202018];
 C.D. Lu and M.Z. Yang, Eur. Phys. J. {\bf C 28}, 515 (2003) [arXiv: hep-ph/0212373].



\bibitem{ioffe 1}B.L. Ioffe and A.V. Smilga, Phys. Lett. {\bf B114},
 353 (1982); Nucl. Phys. {\bf B 216}, 373 (1983).

\bibitem{nesterenko}V.A. Nesterenko and A.V. Radyushkin, Phys. Lett.
{\bf B115}, 410 (1982).


\bibitem{weak decays of QCDSR 1}P. Ball, V.M. Braun and H.G.
Dosch, Phys. Rev. {\bf D 44}, 3567 (1991).

\bibitem{weak decays of QCDSR 2}For a review of QCD sum rules applicaions
to weak decays of heavy mesons, see A. Khodjamirian and R. Ruckl,
Adv. Ser. Direct. High Energy Phys. {\bf 15}, 345 (1998) [arXiv:
hep-ph/9801443].

\bibitem{coupling constants of the strong interactions}V.L. Eletsky, B.L. Ioffe and
 Ya.I. Kogan, Phys. Lett. {\bf B 122}, 423 (1983).

\bibitem{Chernyak}V.L. Chernyak and A.R. Zhitnitsky, Phys. Rep. {\bf 112},
173 (1984);V.L. Chernyak, A.A. Ogloblin and I.R. Zhitnitsky, Z.
Phys. {\bf C 42},  569 (1989); P. Ball, V. M. Braun, Y. Koike and K.
Tanaka, Nucl. Phys. {\bf B 529}, 323 (1998) [arXiv: hep-ph/9802299];
P. Ball, V. M. Braun, Nucl. Phys. {\bf B 529}, 323 (1998) [arXiv:
hep-ph/9810475]; P. Ball, JHEP {\bf{9901}}, 010 (1999) [arXiv:
hep-ph/9812375]; H.Y. Cheng, C.K. Chua  and  K.C. Yang, Phys. Rev.
{\bf D 73}, 014017 (2006) [arXiv: hep-ph/0508104]; C.D.
L$\ddot{\rm{u}}$, Y.M. Wang and H. Zou, Phys. Rev. {\bf D 75},
056001 (2007) [arXiv: hep-ph/0612210].

\bibitem{p. ball}P. Ball, Phys. Rev. {\bf D 48}, 3190 (1993) [arXiv: hep-ph/9305267].






\bibitem{wirbel}M. Wirbel, B. Stech and M. Bauer, Z. Phys. {\bf
C 29}, 637 (1985).

\bibitem{kagan} Talk given by A. Kagan at 4th International Workshop on the CKM
 Unitarity Triangle (CKM 2006), Nagoya, Japan, 12-16 Dec 2006.


\bibitem{kiselev}V.V. Kiselev, A.K. Likhoded and A.I. Onishchenko,
 Nucl. Phys. {\bf B 569},  473  (2000) [arXiv: hep-ph/9905359].


\bibitem{Coulomb corrections}V.V. Kiselev, A.E. Kovalsky and A.K.
Likhoded, Nucl. Phys. {\bf B 585},  353  (2000) [arXiv:
hep-ph/0002127].



\bibitem{t.m. aliev}T.M. Aliev and M. Savci, Eur. Phys. J.  {\bf C 47},  413 (2006)
[arXiv: hep-ph/0601267].



\bibitem{yangmz}D.S. Du, J.W. Li and M.Z. Yang, Eur. Phys. J.  {\bf C 37},  173 (2004)
[arXiv: hep-ph/0308259].


\bibitem{yangmz more}M.Z. Yang, Phys. Rev. {\bf{D 73}},  034027 (2006) [arXiv: hep-ph/0509103];
{\bf{D 73}},  079901 (2006) (E).

\bibitem{ioffe 2}B.L. Ioffe, Prog. Part. Nucl. Phys. {\bf{56}}, 232
 (2006)  [arXiv: hep-ph/0502148].

\bibitem{PDG}W.M. Yao {\it et al.,} J. Phys. {\bf{G 33}}, 1 (2006).

\bibitem{korner}M.A. Ivanov, J.G. K$\ddot{\rm{o}}$rner and P. Santorelli,
Phys. Rev. {\bf{D 73}},  054024 (2006) [arXiv: hep-ph/0602050].

\bibitem{ali khan}A. Ali Khan, V. Braun, T. Burch, M. Gockeler, G. Lacagnina,
A. Schafer, G. Schierholz, [arXiv: hep-lat/0701015].


\bibitem{choi} H.-M. Choi, Phys. Rev. {\bf{D 75}}, 073016 (2007) [arXiv: hep-ph/0701263].


\bibitem{ebert}D. Ebert, R.N. Faustov and V.O. Galkin, Phys. Lett. {\bf{B 635}},
93 (2006) [arXiv: hep-ph/0602110].

\bibitem{aubin}C. Aubin {\it et al.,} Phys. Rev. Lett. {\bf{95}}, 122002 (2005)
[arXiv: hep-lat/0506030].


\bibitem{chiu}T. W. Chiu, T.H. Hsieh, J.Y. Lee, P.H.  Liu and H.J. Chang,
Phys. Lett. {\bf{B 624}}, 31 (2005) [arXiv: hep-ph/0506266].


\bibitem{UKQCD}L. Lellouch and C.-J. Lin (UKQCD Collaboration), Phys. Rev.
{\bf{D 64}}, 094501 (2001) [arXiv: hep-ph/0011086].


\bibitem{bordes}J. Bordes, J. Pe$\tilde{\rm{n}}$arrocha, and K. Schilcher,
JHEP {\bf{0511}}, 014 (2005) [arXiv: hep-ph/0507241].


\bibitem{narison}S. Narison, arXiv: hep-ph/0202200.


\bibitem{rolf}A. Juttner and J. Rolf, Invited talk at 2nd Workshop
on the CKM Unitarity Triangle, Durham, England, 5-9 April, 2003
[arXiv: hep-ph/0306299].


\bibitem{khodjamirian}A.  Khodjamirian, Invited talk at 2nd Workshop
 on the CKM Unitarity Triangle, Durham, England, 5-9 April, 2003
[arXiv: hep-ph/0306253].


\bibitem{becirevic 1}D. Becirevic, Nucl. Phys. Proc. Suppl. {\bf{94}}, 337
(2001) [arXiv: hep-lat/0011075].


\bibitem{becirevic 2}D. Becirevic, P. Boucaud, J.P. Leroy, V. Lubicz,
G. Martinelli, F. Mescia and F. Rapuano, Phys. Rev. {\bf{D 60}},
074501 (1999) [arXiv: hep-lat/9811003].



\bibitem{CLEO D}CLEO Collaboration, M. Artuso {\it et al.,} Phys. Rev. Lett.
{\bf{95}}, 251801 (2005) [arXiv: hep-ex/0508057].


\bibitem{BABAR Ds}BABAR Collaboration, B. Aubert {\it et al.,} Phys. Rev. Lett. {\bf{98}}, 141801
(2007)  [arXiv: hep-ex/0607094].

\bibitem{k.c. bowler} UKQCD Collaboration, K. C. Bowler {\it et
al.,} Nucl. Phys. {\bf{B 619}}, 507 (2001) [arXiv:
 hep-ph/0007020].


\bibitem{CLEO Ds1}CLEO Collaboration, M. Artuso {\it et al.,}
arXiv: 0704.0629 [hep-ex].


\bibitem{CLEO Ds2}CLEO Collaboration, T. K. Pedlar {\it et al.,}
arXiv: 0704.0437 [hep-ex].


\bibitem{kiselev Bc} V.V. Kiselev, arXiv: hep-ph/021102.




\bibitem{dosch}H.G. Dosch, E.M. Ferreira,  F.S. Navarra and  M.
 Nielsen,  Phys. Rev. {\bf{D 65}}, 114002 (2002) [arXiv:
 hep-ph/0203225].

\bibitem{matheus}R.D. Matheus, F.S. Navarra, M. Nielsen and R. Rodrigues da
 Silva,  Phys. Lett. {\bf{B 541}}, 265  (2002) [arXiv:
 hep-ph/0206198].

\bibitem{bracco}M.E. Bracco, M. Chiapparini, F.S. Navarra and M.
 Nielsen, Phys. Lett. {\bf{B 605}}, 326 (2005) [arXiv: hep-ph/0410071].

\bibitem{navarra}F.S. Navarra, Marina Nielsen, M.E. Bracco,
  M. Chiapparini and C.L. Schat, Phys. Lett. {\bf{B 489}}, 319  (2000)
 [arXiv: hep-ph/0005026].

\bibitem{Colangelo}P. Colangelo, G. Nardulli and N. Paver, Z. Phys. {\bf{C 57}},
43 (1993).

\bibitem{yangkc}K.C. Yang and W.Y.P. Hwang, Z. Phys. {\bf{C 73}}, 275
 (1997).



\bibitem{khodjamirian heavy falvors}A. Khodjamirian and R. Ruckl,
 Adv. Ser. Direct. High Energy Phys. {\bf{B 15}}, 345 (1998) [arXiv:
 hep-ph/9801443].


\bibitem{burford}UKQCD Collaboration, D.R. Burford {\it et al.,}
 Nucl. Phys. {\bf{B 447}}, 425 (1995) [arXiv: hep-lat/9503002].

\bibitem{h.y. cheng}Y.H. Chen, H.Y. Cheng, B. Tseng and K.C. Yang,
 Phys. Rev. {\bf{D 60}}, 094014 (1999) [arXiv: hep-ph/9903453].


\bibitem{ISGW}N. Isgur D. Scora, B. Grinstein and M.B. Wise,
Phys. Rev. {\bf{D 39}}, 799 (1989).


\bibitem{Kiselev prd}V.V. Kiselev  and A.V. Tkabladze,
Phys. Rev. {\bf{D 48}}, 5208 (1993).



\bibitem{Wangetal} Y.M. Wang, {\it et al}., in preparation.




\bibitem{landau equation}L.D. Landau,  Nucl. Phys.  {\bf{B 13}},
 181 (1959).

\bibitem{S matrix}For a review, see T.J. Eden, P.V. Landshoff, D.I. Olive and J.C.
 Polkinghorne, ``The Analytic S-Matrix", (Cambridge University
 Press, Cambridge, England, 1966).


 \bibitem{reinders}L.J. Reinders, H. Rubinstein and S. Yazaki, Phys. Rep. {\bf{127}}, 1 (1985).



\bibitem{reinders P} L.J. Reinders, H.R. Rubinstein, S. Yazaki, Phys. Lett. {\bf{B 97}}, 257
(1980);  {\bf{B 100}}, 519 (1981)(E).

\bibitem{belyaev}V.M. Belyaev, V.M. Braun, A. Khodjamirian and R. R\"{u}ckl,
Phys. Rev. {\bf{D 51}}, 6177 (1994) [arXiv: hep-ph/9410280].

\bibitem{bediaga}I. Bediaga and M. Nielsen, Phys. Rev. {\bf{D 68}}, 036001 (2003) [arXiv: hep-ph/0304193].


\bibitem{reinders V}L.J. Reinders, S. Yazaki  and H.R. Rubinstein, Phys. Lett. {\bf{B 103}},
63 (1981).


\end{thebibliography}
\end{document}